\documentclass[a4paper,fleqn]{cas-dc}

\everymath=\expandafter{\the\everymath\displaystyle}
\usepackage{amsmath}
\usepackage{tikz}
\usepackage{caption}
\usepackage{subcaption}
\usepackage{amsfonts}
\usepackage{amssymb}
\usepackage{bm}
\usepackage{pgfplots}
\usepackage{float}
\usepackage{svg}
\usepackage{graphicx}
\usepackage{rotating}
\usepackage{threeparttable}
\usepackage{tabularx}
\usepackage{awesomebox}
\usepackage{enumitem}
\usepackage[numbers]{natbib}

\def\tsc#1{\csdef{#1}{\textsc{\lowercase{#1}}\xspace}}
\tsc{WGM}
\tsc{QE}
\definecolor{noteblue}{RGB}{154, 191, 219}
\definecolor{nutshellgray}{RGB}{195, 195, 196}

\newif\ifbluetext
\bluetextfalse  

\newcommand{\blue}[1]{%
  \ifbluetext
    \ifmmode
      {\color{blue}#1}
    \else
      \textcolor{blue}{#1}
    \fi
  \else
    #1%
  \fi
}

\newcommand\blfootnote[1]{%
  \begingroup
  \renewcommand\thefootnote{}\footnote{#1}%
  \addtocounter{footnote}{-1}%
  \endgroup
}

\begin{document}
\let\WriteBookmarks\relax
\def\floatpagepagefraction{1}
\def\textpagefraction{.001}

\shorttitle{}    

\shortauthors{}  

\title [mode = title]{A tutorial overview of model predictive control for continuous crystallization: current possibilities and future perspectives}  



%

\author[1]{Collin R. Johnson}






\affiliation[1]{organization={Chair of Process Automation Systems, TU Dortmund University},
            addressline={Emil-Figge-Str. 70}, 
            city={Dortmund},
            postcode={44227}, 
            country={Germany}}

\author[2]{Kerstin Wohlgemuth}
\affiliation[2]{organization={Laboratory of Plant and Process Design, TU Dortmund University},
            addressline={Emil-Figge-Str. 70}, 
            city={Dortmund},
            postcode={44227}, 
            country={Germany}}





\author[1]{Sergio Lucia}




\begin{abstract}
\protect\blue{Continuous crystallization processes require advanced control strategies to ensure consistent product quality, yet deploying optimization-based controllers such as model predictive control remains challenging. Combining spatially distributed crystallizer models with detailed particle size distributions leads to computationally demanding problems that are difficult to solve in real-time. This tutorial provides a comprehensive overview of how to address this challenge. Topics include numerical methods for solving population balance equations, modeling of crystallizers, and data-driven surrogate modeling. We show how these elements combine within a model predictive control framework to enable real-time control of particle size distributions. Two case studies illustrate the complete workflow: a well-mixed crystallizer that allows comparison with established methods, and a spatially distributed plug-flow crystallizer that demonstrates application to more complex systems. Readers gain a practical roadmap for implementing model predictive control in continuous crystallization, supported by open-source code and interactive examples. The tutorial concludes by outlining open challenges and emerging opportunities in the field.}
\end{abstract}





\maketitle

\blfootnote{This work was funded by the Deutsche Forschungsgemeinschaft (DFG, German Research Foundation) – 504676854 – within the Priority Program “SPP 2364: Autonomous processes in particle technology”.}

\section{Introduction}\label{section:Introduction}

Crystallization is a fundamental separation process widely applied in various industries, ranging from sugar production \citep{moralesCrystallizationProcessSugar2024} to pharmaceuticals \citep{chenPharmaceuticalCrystallization2011}, where achieving product purity and quality is critical. In the pharmaceutical sector, stringent demands on product quality require precise control over particle properties, such as particle size distribution, directly affecting drug effectiveness \citep{shekunovParticleSizeAnalysis2007}. Traditionally, batch crystallization processes are favored when a specific particle size distribution is needed. Although continuous crystallizers have long been employed for petrochemicals, batch processes have predominated in the pharmaceutical industry \citep{yazdanpanahHandbookContinuousCrystallization2020}.

In recent years, continuous crystallization has been recognized as beneficial in many areas, including pharmaceutical processes \citep{plumbContinuousProcessingPharmaceutical2005}. Compared to batch methods, continuous processes offer advantages, such as reduced batch-to-batch variability and improved reproducibility. In addition, they reduce both operational and capital costs \citep{chenPharmaceuticalCrystallization2011}.

Effective process control is essential to fully leverage the advantages of continuous crystallization. Model predictive control (MPC) is a widely used advanced control technique capable of handling nonlinear systems, constraints, and multiple control objectives \citep{rawlingsModelPredictiveControl2017}. MPC has been successfully applied to both batch and continuous crystallization processes, yet most implementations rely on moment-based methods to model the particle size distribution \citep{baloghlaszloModelPredictiveControl2023, kwonModelingControlCrystal2014, moldovanyiModelPredictiveControl2005, yangAdvancedControlApproaches2015, caoRealtimeFeasibleMultiobjective2017}. Although numerically simple, these methods struggle to accurately represent full particle size distributions. Furthermore, many applications found in the literature are limited to lumped parameter models that neglect spatial variations. However, many continuous crystallization systems, such as slug-flow crystallizers \cite{jiangContinuousFlowTubularCrystallization2014} and continuous oscillatory baffled crystallizers \cite{penaProcessIntensificationContinuous2017}, require detailed spatial modeling. The combination of spatially distributed crystallizer models with population balance equations (PBE) often leads to complex partial differential equations (PDEs), which pose significant computational challenges for optimization-based methods like MPC.

\begin{figure*}
    \centering
    \includegraphics[width=\textwidth]{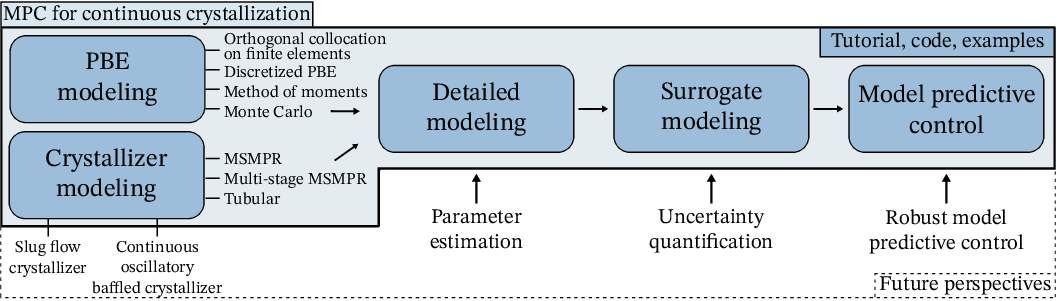}
    \caption{Outline of the presented work. PBE modeling \ref{section:SolutionMethods} and crystallization modeling \ref{section:Models} lead to a detailed model. The detailed model is approximated by a data-based surrogate model \ref{section:ControlOriented} and is used in MPC \ref{section:MPC}. This work provides a tutorial overview, code and examples for all the blue boxes in the gray area. The topics in white area are discussed as future perspectives.}
    \label{figure:Overview}
\end{figure*}

Recent advances in machine learning have made these challenges manageable. Data-driven methods offer a promising solution by enabling optimization-based control even when detailed models for particle size distribution and crystallization dynamics are used. Surrogate models can approximate complex first-principle models while maintaining computational efficiency, making real-time optimization feasible. However, most existing approaches only track auxiliary states, e.g. the concentration of the liquid phase, indirectly controlling the particle size distribution rather than explicitly regulating its key characteristics \citep{limaDevelopmentRecurrentNeural2022, rohaniModelingControlContinuous1999}. 

Despite these advances, it has not yet been computationally feasible to develop MPC schemes that incorporate both a detailed particle size distribution model and a spatially distributed crystallizer model. As a result, there is currently no comprehensive overview of dynamic modeling techniques that address both aspects. This work aims to address this gap by providing a tutorial overview of the most suitable dynamic modeling strategies for particle size distribution and spatially distributed crystallizers. We outline the steps required to develop machine learning-based surrogate models and demonstrate how they can be leveraged to construct computationally tractable MPC controllers enabling advanced process control for continuous crystallization processes. With this tutorial, we aim to foster stronger collaboration between the communities of advanced process control and continuous crystallization.
 
Figure \ref{figure:Overview} presents an outline of the presented workflow. \blue{We present in detail the steps in the gray area and briefly discuss the ideas in the white area.} This paper is structured so that the steps of the workflow are described from left to right. \blue{First, in Section \ref{section:LiteratureReview} existing literature is reviewed, and our contribution is situated within the field.} In Sections \ref{section:SolutionMethods} and \ref{section:Models} we present the modeling and solution techniques for continuous crystallization systems, resulting in a detailed model. Sections \ref{section:ControlOriented} and \ref{section:MPC} present the transformation to a control-oriented surrogate model and the subsequent application in MPC. Section \ref{section:FuturePerspectives} discusses future perspectives in the field based on recent advances, as seen in the lower part of Figure \ref{figure:Overview}. Finally, we conclude in Section \ref{section:Conclusion}.

We provide the code for the solution methods of the PBE, the crystallizer models, the surrogate modeling, and the code showing the application in MPC in our GitHub repository\footnote{\href{https://github.com/collinj2812/do-crystal}{https://github.com/collinj2812/do-crystal}}. In addition, we present the different solution methods as well as models in an interactive dashboard\footnote{\href{https://pbe-solutions.streamlit.app/}{https://pbe-solutions.streamlit.app/}}.

\section{\blue{Literature review}}\label{section:LiteratureReview}

\blue{Controlling continuous crystallizers is a difficult task. The parameters in continuous crystallizers are usually distributed in space and in particle size. This leads to the resulting models being more complex. At the same time, crystallization processes are nonlinear and usually involve different time scales. These characteristics pose challenges for model-based control, which requires models that are both accurate and computationally tractable. In order to operate continuous crystallization processes with model-based control, various sub-areas must be combined. Process engineers design and operate the crystallizer. To obtain an accurate model, numerical solutions for PBEs and fluid dynamics must be found. Finally, model-based control algorithms must be developed and implemented, and machine learning can be used to bring everything together.}

\blue{Table~\ref{tb:ReviewComparisons} presents the coverage across major reviews, revealing that while individual topics and pairwise combinations are addressed, no existing review spans the full intersection required for MPC for spatially-distributed crystallization. There are several review articles in the literature that combine some of these tasks, but not all of them. Some reviews discuss the combination of PBE modeling and fluid dynamic modeling to obtain accurate and complex models \cite{omarCrystalPopulationBalance2017, shieaNumericalMethodsSolution2020}. However, integration into a control algorithm is missing. Reviews exist for the control of crystallization systems \cite{braatzAdvancedControlCrystallization2002,  nagyAdvancesNewDirections2012, orehekContinuousCrystallizationProcesses2021}, but these assume well-mixed systems, i.e. explicitly not systems that are spatially distributed. Furthermore, novel continuous crystallizers continue to be developed \cite{jiangDesignsContinuousflowPharmaceutical2019}. Modeling and control of spatially-distributed continuous systems is not well developed. Reviews exist on the application of AI in crystallization \cite{xiourasApplicationsArtificialIntelligence2022, wuTutorialReviewMachine2025, limaApplicationsMachineLearning2025}. However, here also, the application to spatially distributed systems is not yet represented.}
\newcommand\x{60}
\begin{table*}
    \centering
    \begin{threeparttable}
\caption{\blue{Comparative table for existing review articles.}}\label{tb:ReviewComparisons}
\begin{tabular}{@{}l*{11}{p{0.6cm}}@{}}
\toprule
\blue{Topic} &
\rotatebox[origin=lb]{\x}{\blue{\citet{braatzAdvancedControlCrystallization2002}}} &
\rotatebox[origin=lb]{\x}{\blue{\citet{yuRecentAdvancesCrystallization2007}}} &
\rotatebox[origin=lb]{\x}{\blue{\citet{nagyAdvancesNewDirections2012}}} &
\rotatebox[origin=lb]{\x}{\blue{\citet{jiangDesignsContinuousflowPharmaceutical2019}}} &
\rotatebox[origin=lb]{\x}{\blue{\citet{orehekContinuousCrystallizationProcesses2021}}} &
\rotatebox[origin=lb]{\x}{\blue{\citet{xiourasApplicationsArtificialIntelligence2022}}} &
\rotatebox[origin=lb]{\x}{\blue{\citet{wuTutorialReviewMachine2025}}} &
\rotatebox[origin=lb]{\x}{\blue{\citet{limaApplicationsMachineLearning2025}}} &
\rotatebox[origin=lb]{\x}{\blue{\citet{omarCrystalPopulationBalance2017}}} &
\rotatebox[origin=lb]{\x}{\blue{\citet{shieaNumericalMethodsSolution2020}}} &
\rotatebox[origin=lb]{\x}{\blue{This tutorial}}
\\
\midrule
\blue{Spatial + full PSD} & \blue{X} & \blue{X} & \blue{X} & \blue{X} & \blue{O} & \blue{O} & \blue{X} & \blue{X} & \blue{\checkmark} & \blue{\checkmark} & \blue{\checkmark} \\
\blue{MPC / advanced control} & \blue{\checkmark} & \blue{O} & \blue{\checkmark} & \blue{X} & \blue{\checkmark} & \blue{O} & \blue{\checkmark} & \blue{\checkmark} & \blue{X} & \blue{X} & \blue{\checkmark} \\
\blue{ML / surrogate models} & \blue{X} & \blue{X} & \blue{X} & \blue{X} & \blue{X} & \blue{\checkmark} & \blue{\checkmark} & \blue{\checkmark} & \blue{X} & \blue{X} & \blue{\checkmark} \\
\blue{Continuous crystallizers} & \blue{O} & \blue{X} & \blue{\checkmark} & \blue{\checkmark} & \blue{\checkmark} & \blue{O} & \blue{X} & \blue{O} & \blue{X} & \blue{X} & \blue{\checkmark} \\
\blue{PBE solution methods} & \blue{\checkmark} & \blue{O} & \blue{O} & \blue{X} & \blue{O} & \blue{X} & \blue{X} & \blue{O} & \blue{\checkmark} & \blue{\checkmark} & \blue{\checkmark} \\
\blue{Crystallization-specific} & \blue{\checkmark} & \blue{\checkmark} & \blue{\checkmark} & \blue{\checkmark} & \blue{\checkmark} & \blue{\checkmark} & \blue{X} & \blue{\checkmark} & \blue{\checkmark} & \blue{X} & \blue{\checkmark} \\
\bottomrule
\end{tabular}
    \begin{tablenotes}
        \footnotesize
        \item \blue{\checkmark = addressed in depth, O = mentioned/referenced, X = not covered}
    \end{tablenotes}
    \end{threeparttable}
\end{table*}
\blue{The use of machine learning algorithms to enable MPC has recently gained more attention. The review by \cite{wuTutorialReviewMachine2025} comprehensively demonstrates the use of neural networks in learning-based MPC. However, the applications are rather simple reaction systems in which the states are neither spatially distributed nor part of a particle size distribution. \cite{xiourasApplicationsArtificialIntelligence2022} demonstrate a wide range of applications for crystallization. The examples in this review focus on structure prediction, property estimation, and process control. However, the application to control is only shown at a high level. Spatially and particulate distributed systems are only mentioned as a future perspective. Recent work by \cite{limaApplicationsMachineLearning2025} bridges the gap between machine learning and crystallization control. Hybrid methods, physics-informed methods, and reinforcement methods are presented. Nevertheless, the application to spatially distributed continuous crystallizers is not addressed, as the examples here are not spatially distributed, but rather well-mixed systems.}

\blue{The pattern across these reviews is consistent: the methodological tools for ML-based MPC exist, the crystallization application domain is well-mapped, yet spatially-distributed crystallization systems remain deferred to future work. This tutorial addresses that intersection.}

\blue{In the following, we present some trends in the literature for the four main topics that we cover in this tutorial overview. We will present representative studies for the most important PBE solution methods, crystallizer models, the use of surrogate models in crystallization, and the application of MPC in continuous crystallization. Table~\ref{tb:PBELiterature} presents some research work on different PBE solution methods in crystallization.}
\begin{table*}
    \centering
    \begin{threeparttable}
    \caption{\blue{Representative studies on PBE solution methods for crystallization modeling (our tutorial in Section~\ref{section:SolutionMethods}).}}
    \label{tb:PBELiterature}
    \centering
    \begin{tabularx}{\textwidth}{@{}llllllll@{}}
        \toprule
        \blue{Reference} & \blue{Year} & \blue{Method} & \blue{System} & \blue{\makecell[c]{Dimensions in \\ PBE}} & \blue{\makecell[c]{Dimensions in \\ cont. phase}} & \blue{Phenomena} \\
        \midrule
        \blue{\citet{randolphTransientSteadyState1962a}} & \blue{1962} & \blue{Analytical/FD} & \blue{MSMPR} & \blue{1D} & \blue{0D} & \blue{N, G} \\
        \blue{\citet{mcgrawDescriptionAerosolDynamics1997}} & \blue{1997} & \blue{QMOM} & \blue{Aerosol} & \blue{1D} & \blue{0D} & \blue{N, G} \\
        \blue{\citet{gunawanHighResolutionAlgorithms2004}} & \blue{2004} & \blue{HR-FVM} & \blue{Batch/MSMPR} & \blue{1--2D} & \blue{0D} & \blue{N, G} \\
        \blue{\citet{briesenSimulationCrystalSize2006}} & \blue{2006} & \blue{2D-FD} & \blue{Batch} & \blue{2D} & \blue{0D} & \blue{G} \\
        \blue{\citet{aamirCombinedQuadratureMethod2009}} & \blue{2009} & \blue{QMOM-MOC} & \blue{Batch} & \blue{1D} & \blue{0D} & \blue{N, G} \\
        \blue{\citet{szilagyiGraphicalProcessingUnit2016}} & \blue{2016} & \blue{GPU HR-FVM} & \blue{Batch} & \blue{1--2D} & \blue{0D} & \blue{N, G} \\
        \blue{\citet{sulttanCouplingCFDPopulation2019}} & \blue{2019} & \blue{HR-FVM} & \blue{MSMPR-helical tubular} & \blue{1D} & \blue{1D} & \blue{N, G} \\
        \blue{\citet{huangNumericalInvestigationCrystal2024}} & \blue{2024} & \blue{QMOM} & \blue{COBC} & \blue{1D} & \blue{3D} & \blue{N, G, A, B} \\
        \bottomrule
    \end{tabularx}
    \begin{tablenotes}
        \footnotesize
        \item \blue{Methods: FD=finite differences, QMOM=quadrature method of moments, HR-FVM=high resolution finite volume method, MOC=method of characteristics, GPU=graphics processing unit.}
        \item \blue{Phenomena: N=nucleation, G=growth, A=agglomeration, B=breakage.}
    \end{tablenotes}
    \end{threeparttable}
\end{table*}
\blue{PBE solution methods have evolved over the years. Initially, analytical solutions were sought for the PBE as in \cite{randolphTransientSteadyState1962a}. However, these are very limited and only available for certain formulations of the PBE, and therefore numerical methods were applied and developed. Due to the advancement of fast computing, a trend can be seen over the years from computationally less demanding moment-based methods \cite{mcgrawDescriptionAerosolDynamics1997} to more advanced and computationally demanding high-resolution finite volume methods \cite{gunawanHighResolutionAlgorithms2004, szilagyiGraphicalProcessingUnit2016}. The subsequent use in model-based control significantly increases the complexity of the overall problem dramatically. Coupling with continuous phase is still viable only for well-mixed systems. The direct use of models that are discretized in multiple dimensions in optimization remains very challenging to this day.}

\blue{Table~\ref{tb:CrystallizerLiterature} presents some representative references modeling the continuous phase of different continuous crystallizers.}
\begin{table*}
    \centering
    \begin{threeparttable}
    \caption{\blue{Representative studies on the modeling of continuous crystallizers (our tutorial in Section~\ref{section:Models}).}}
    \label{tb:CrystallizerLiterature}
    \centering
    \begin{tabularx}{\textwidth}{@{}llllllll@{}}
        \toprule
        \blue{Reference} & \blue{Year} & \blue{Crystallizer} & \blue{\makecell[c]{PBE solution \\ method}} & \blue{Phenomena} & \blue{Chemical system} \\
        \midrule
        \blue{\citet{randolphMixedSuspensionMixed1965}} & \blue{1965} & \blue{MSMPR} & \blue{Analytical} & \blue{N, G} & \blue{Not specified} \\
        \blue{\citet{suPharmaceuticalCrystallisationProcesses2015}} & \blue{2015} & \blue{MSMPR cascade} & \blue{SMOM} & \blue{N, G} & \blue{Paracetamol in acetone-water} \\
        \blue{\citet{sang-ilkwonCrystalShapeSize2014}} & \blue{2014} & \blue{Plug flow} & \blue{SMOM} & \blue{G} & \blue{Lysozyme} \\
        \blue{\citet{mcgloneOscillatoryFlowReactors2015}} & \blue{2015} & \blue{Continuous oscillatory flow} & \blue{Various} & \blue{Various} & \blue{Various} \\
        \blue{\citet{kufnerModelingContinuousSlug2023}} & \blue{2023} & \blue{Slug flow} & \blue{HR-FVM} & \blue{G, A} & \blue{L-alanine in water} \\
        \bottomrule
    \end{tabularx}
    \begin{tablenotes}
        \footnotesize
        \item \blue{Methods: SMOM=standard method of moments, HR-FVM=high resolution finite volume method.}
        \item \blue{Phenomena: N=nucleation, G=growth, A=agglomeration.}
    \end{tablenotes}
    \end{threeparttable}
\end{table*}
\blue{Modeling of the continuous phase of crystallizers (usually concentration and temperature) consists of mass and energy balances, which makes it conceptually straightforward. For well-mixed MSMPR systems \cite{randolphMixedSuspensionMixed1965}, the assumption of perfect mixing eliminates spatial gradients, reducing the model to ordinary differential equations in time. Spatially distributed crystallizers such as tubular plug-flow systems \cite{sang-ilkwonCrystalShapeSize2014} lead to partial differential equation systems also for the continuous phase, but the underlying principles remain mass and energy conservation. More specialized crystallizers, such as continuous oscillatory baffled crystallizers \cite{mcgloneOscillatoryFlowReactors2015} and slug flow crystallizers \cite{kufnerModelingContinuousSlug2023} present additional challenges as their unique hydrodynamic behavior requires adapted or entirely new modeling approaches.}

\blue{Recently, machine learning has proven useful for combining complex, accurate models for crystallization with MPC using surrogate models \cite{zhengPredictiveControlBatch2022, wuPhysicsinformedMachineLearning2023, johnsonMultistageModelPredictive2026}. Several new studies have already been conducted that use surrogate models to reduce the complexity of the model. Table~\ref{tb:SurrogateModelsLiterature} lists several studies on the application of surrogate models in crystallization.}
\begin{table*}
    \caption{\blue{Representative studies for surrogate modeling approaches applied to crystallization processes (our tutorial in Section~\ref{section:ControlOriented}).}}
    \label{tb:SurrogateModelsLiterature}
    \centering
    \begin{threeparttable}
    \begin{tabularx}{\textwidth}{@{}lllllllll@{}}
        \toprule
        \blue{Reference} & \blue{Year} & \blue{Surrogate} & \blue{Data source} & \blue{Crystallizer} & \blue{MPC} \\
        \midrule
        \blue{\citet{rohaniModelingControlContinuous1999}} & \blue{1999} & \blue{Recurrent neural networks/neural networks} & \blue{Simulation} & \blue{MSMPR} & \blue{Yes} \\
        \blue{\citet{sanzidaPolynomialChaosExpansion2014}} & \blue{2014} & \blue{Polynomial chaos expansion} & \blue{PBE} & \blue{Batch} & \blue{No$^*$} \\
        \blue{\citet{makrygiorgosSurrogateModelingFast2020}} & \blue{2020} & \blue{Sparse polynomial chaos-Kriging} & \blue{2D-PBE} & \blue{Batch} & \blue{No} \\
        \blue{\citet{sitapureMultiscaleCFDModeling2021}} & \blue{2021} & \blue{Neural networks} & \blue{kinetic MC} & \blue{Slug flow} & \blue{No$^*$} \\
        \blue{\citet{zhengMachineLearningModeling2022}} & \blue{2022} & \blue{Auto-encoder recurrent neural networks} & \blue{PBE} & \blue{Batch} & \blue{Yes} \\
        \blue{\citet{songSurrogateModelingSensitivityBased2025}} & \blue{2025} & \blue{Gaussian process regression} & \blue{Experimental} & \blue{Batch} & \blue{No} \\
        \blue{\citet{johnsonMultistageModelPredictive2026}} & \blue{2026} & \blue{Conformalized quantile regression/Bayesian last layer} & \blue{PBE} & \blue{Slug flow} & \blue{Yes} \\
        \bottomrule
    \end{tabularx}
    \begin{tablenotes}
        \footnotesize
        \item \blue{$^*$used in optimization, not closed-loop control.}
        \item \blue{MC=Monte Carlo.}
    \end{tablenotes}
    \end{threeparttable}
\end{table*}
\blue{Surrogate models are not only used for control. In some cases, surrogate models are used to calculate parameters such as growth rate \cite{sitapureMultiscaleCFDModeling2021}. Other studies use surrogate models for one-time optimization of the process but have not yet closed the control loop \cite{sanzidaPolynomialChaosExpansion2014, sitapureMultiscaleCFDModeling2021}.}

\blue{Subsequently, Table~\ref{tb:MPCLiterature} lists papers from the literature that apply MPC to crystallization systems.}
\begin{table*}
    \centering
    \begin{threeparttable}
    \caption{\blue{Representative MPC studies for crystallization processes (our tutorial in Section~\ref{section:MPC}).}}
    \label{tb:MPCLiterature}
    \centering
    \begin{tabularx}{\textwidth}{@{}llllllllll@{}}
        \toprule
    \blue{Reference} & \blue{Year} & \blue{\makecell[c]{Crystallizer \\ type}} & \blue{\makecell[c]{Model \\ type}} & \blue{\makecell[c]{PSD \\ representation}} & \blue{\makecell[c]{Spatial \\ treatment}} & \blue{\makecell[c]{Comp. \\ time}} & \blue{\makecell[c]{Real-time \\ feasible}} \\
        \midrule
        \blue{\citet{rohaniModelingControlContinuous1999a}} & \blue{1999} & \blue{Continuous cooling} & \blue{NN surrogate} & \blue{None} & \blue{None} & \blue{3--18 s} & \blue{Yes} \\
        \blue{\citet{moldovanyiModelPredictiveControl2005}} & \blue{2005} & \blue{MSMPR} & \blue{PBE} & \blue{Moments} & \blue{None} & \blue{NR} & \blue{---$^*$} \\
        \blue{\citet{shiPredictiveControlParticle2006}} & \blue{2006} & \blue{Continuous/batch} & \blue{PBE} & \blue{Moments} & \blue{None} & \blue{NR} & \blue{---$^*$} \\
        \blue{\citet{yangAdvancedControlApproaches2015}} & \blue{2015} & \blue{MSMPR cascade} & \blue{PBE} & \blue{Moments} & \blue{None} & \blue{$<$2 min} & \blue{Yes} \\
        \blue{\citet{szilagyiChordLengthDistribution2018}} & \blue{2018} & \blue{Batch cooling} & \blue{PBE} & \blue{Full PSD} & \blue{None} & \blue{160 s} & \blue{Yes} \\
        \blue{\citet{zhengPredictiveControlBatch2022}} & \blue{2022} & \blue{Batch cooling} & \blue{RNN surrogate} & \blue{Full PSD} & \blue{None} & \blue{25--1000 s} & \blue{Yes} \\
        \blue{\citet{wuPhysicsinformedMachineLearning2023}} & \blue{2023} & \blue{Batch cooling} & \blue{PIRNN surrogate} & \blue{Moments} & \blue{None} & \blue{NR} & \blue{---$^*$} \\
        \blue{\citet{johnsonMultistageModelPredictive2026}} & \blue{2026} & \blue{Slug-flow} & \blue{NN surrogate} & \blue{Full PSD} & \blue{Implicit} & \blue{2--4 s} & \blue{Yes} \\
        \bottomrule
    \end{tabularx}
    \begin{tablenotes}
        \footnotesize
        \item \blue{NR=not reported, $^*$therefore check for real-time feasibility not possible.}
        \item \blue{NN=neural network, RNN=recurrent neural network, PIRNN=physics informed neural network.}
    \end{tablenotes}
    \end{threeparttable}
\end{table*}
\blue{The trend that MPC is mostly only used for well-mixed batch or continuous crystallizers is evident. In addition, moment-based methods \cite{moldovanyiModelPredictiveControl2005, shiPredictiveControlParticle2006, yangAdvancedControlApproaches2015, wuPhysicsinformedMachineLearning2023} are mostly used for the PBE to keep the complexity of the model even lower.}

\blue{Taken together, these four aspects of MPC for continuous crystallization reveal a persistent trade-off: full PSD resolution and spatial distribution have been achieved separately, but their combination within real-time MPC remains open for further research. With this tutorial, we present all the necessary tools to achieve this goal.}

\section{Population balance equation and solution methods}\label{section:SolutionMethods}

\begin{figure}
    \centering
    \includegraphics{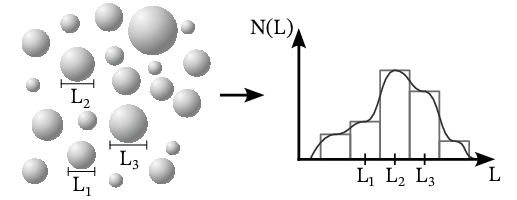}
    \caption{Illustration of the connection between a population of particles and the function of the particle size distribution. The particle size distribution can be regarded as a histogram over some property of the particles, in this case the length. The number distribution $N(L)$ can be transformed in the number density distribution $n(L)$ by dividing by the width of the classes.}
    \label{figure:SketchPopulation}
\end{figure}

The population balance equation (PBE) tracks the temporal evolution of some population and was first introduced by \citet{hulburtProblemsParticleTechnology1964}. Since its formulation is very general, it can be applied to a broad range of fields. Subsequently, we use terms common to the field of crystallization. Figure~\ref{figure:SketchPopulation} shows the relation between a population of particles and the distribution function which the PBE aims to describe.

As described by \citet{ramkrishnaPopulationBalancesTheory2000}, the particle state space is split into internal and external coordinates. The internal coordinates refer to the characteristics of the particles as such. We use particle length $L$ or volume $V$ as an internal coordinate. In addition, it is possible to track the distribution of the shape of the particles, resulting in a multidimensional distribution. The external coordinates refer to the location of the particles. Following, particles are assumed to be uniformly distributed in space and, therefore, omitting external coordinates within the PBE. For the one-dimensional well-mixed case, the following PBE is obtained:
\begin{equation}\label{eq:PBEGeneral}
    \frac{\partial n(L,t)}{\partial t}+\frac{\partial G(L)n(L,t)}{\partial L}=h(L,t).
\end{equation}
The number density function is represented by $n$ and is a function of time $t$, and the internal coordinate, where we use the particle length $L$. The number density function represents how many particles possess specific characteristics, in this case particle length, per unit volume at a given time. We define $G(L)$ as the growth rate. For the source function $h(L,t)$, containing birth and death terms, nucleation and agglomeration is considered:
\begin{subequations}\label{eq:PBESourceTerms}
    \begin{gather}
        h(L,t) = N+B-D,
    \end{gather}
    \text{with} 
    \begin{align}
        N &= f(L,t), \label{eq:PBESourceTermssubeq:Nucleation} \\
        B &= \frac{1}{2}\int_{0}^{L}\beta(L',L-L')n(L',t)n(L-L',t)dL', \label{eq:PBESourceTermssubeq:Birth} \\
        D &= n(L,t)\int_{0}^{\infty}\beta(L,L')n(L',t)dL', \label{eq:PBESourceTermssubeq:Death}
    \end{align}
\end{subequations}
where $N$ refers to the nucleation rate, $B$ is the birth term associated with agglomeration and $D$ is the corresponding death term. For the calculation of agglomeration, the kernel function $\beta$ must be known. The PBE is therefore complex to solve, combining both integral and partial differential components.

Note that, to keep the notation uncluttered, we omitted the dependence on the properties of the continuous phase in \eqref{eq:PBEGeneral} and \eqref{eq:PBESourceTerms}. To obtain a meaningful model for our system we must couple the PBE with equations describing the continuous phase and functions, e.g., for the nucleation rate $N$, the growth rate $G$, and the kernel $\beta$. These are usually dependent on other process parameters, such as concentration, temperature, and pressure. Coupling the PBE with the equations modeling the continuous phase, i.e. the mass and energy balances, is essential and is covered in Section~\ref{section:Models}. 

This section provides five different solution methods to solve the PBE \eqref{eq:PBEGeneral}. The presented methods and some basic characteristics and differences are shown in Table~\ref{tb:GuideSolutionMethods}. Complexity refers to the complexity of the implementation of the method. Methods also differ in the information they provide. More precisely, the methods provide either the full distribution or only information about the moments. The domain is also bounded in some methods, i.e. there might exist a minimum and a maximum particle length or volume that can be considered. Computational complexity does not necessarily have to correspond to the complexity of the implementation, as is, for example, the case with the MC method, which is simple to implement but results in large simulation times.

\begin{table}
\caption{Comparative guide for solution methods. For complexity and computational expense, $+$ represents lower complexity and less computational expense, respectively.}\label{tb:GuideSolutionMethods}
\begin{tabular*}{\tblwidth}{@{} LLLCC@{} }
\toprule
Method & Information & Domain & Complexity & Comp. expense \\
\midrule
SMOM & Moments & Full & \blue{$++$} & \blue{$++$} \\
QMOM & Moments & Full & \blue{$+$} & \blue{$++$} \\
DPBE & Full PSD & Truncated & \blue{$-$} & \blue{$-$} \\
OCFE & Full PSD & Truncated & \blue{$--$} & \blue{$-$} \\
MC & Full PSD & Full & \blue{$+$} & \blue{$--$} \\
\bottomrule
\end{tabular*}
\end{table}

\subsection{Standard method of moments (SMOM)}\label{subsection:SMOM}

A simple method to solve the PBE is to apply the moment transformation and track the temporal evolution of the moments. The method of moments was first introduced by \citet{hulburtProblemsParticleTechnology1964} and the moment transformation is given by:
\begin{equation}\label{eq:MomentTransformation}
    \mu_k = \int L^k n(L) dL,
\end{equation}
where $\mu_k$ represents the $k$-th moment. Moments can be related to cumulative physical properties of the distribution. For the number distribution, for example, the $0$-th moment corresponds to the total number of particles, whereas the $1$-st moment corresponds to the cumulative length of the distribution when using a length as internal coordinate $L$. Applying the method of moments to the PBE leads to a set of $k$ ordinary differential equations. \blue{The full transformation can be found in \cite{mcgrawDescriptionAerosolDynamics1997}}. However, for the general PBE from \eqref{eq:PBEGeneral}, which contains an arbitrary growth function as well as arbitrary agglomeration kernels, the set of equations is not necessarily closed, e.g. the set of equations describing the evolution of the first $k$ moments might also contain the $k+1$-st moment. In practice, the standard method of moments is only applicable to specific growth functions (e.g. constant growth or a linear function of $L$) and specific agglomeration kernels (e.g. constant or sum kernels). Here, the method of moments is applied to the PBE considering only nucleation and size-independent growth. The resulting differential equation for the $k$-th moment is given by:
\begin{equation}\label{eq:SMOMODEs}
    \frac{d \mu_k}{d t} = kG \mu_{k-1}.
\end{equation}
Assuming nucleation to occur at $L=0$, nucleation can be considered as a boundary condition for the $0$-th moment:
\begin{equation}\label{eq:SMOMNucleationBoundary}
    \frac{d \mu_0}{d t} = N.
\end{equation}
It is worth noting that it is possible to calculate characteristic properties of the number density function using the moments, such as mean diameters and their coefficients of variation. The mean length of the mass distribution $L_{43}$ and its coefficient of variation $\text{CV}$ can be calculated as:
\begin{align}\label{eq:MeanDiameterMomentFunction}
    L_{43} &= \frac{\mu_4}{\mu_3}, \\
    \text{CV} &=\frac{\sqrt{\mu_4-L_{43}^2}}{L_{43}}.
\end{align}

\subsection{Quadrature method of moments (QMOM)}\label{subsection:QMOM}
A method that also tracks the moments of the distribution but overcomes the challenge of a non-closed equation system is the quadrature method of moments \cite{mcgrawDescriptionAerosolDynamics1997}. Here, the moments of the distribution are approximated by a quadrature formula:
\begin{equation}\label{eq:QuadratureApproximationMoments}
    \mu_k = \int L^k n(L) dL=\sum_{i=1}^{N_q} L_i^k w_i.
\end{equation}
The number density function $n(L)$ is approximated at the abscissas $L_i$ using the weights $w_i$:
\begin{equation}\label{eq:QuadratureApproximationDistribution}
    n(L) \approx \sum_{i=1}^{N_q} w_i \delta (L-L_i),
\end{equation}
where $\delta$ represents the Dirac delta function. It is possible to calculate the first $2N_q$ moments using a quadrature formula of order $N_q$. To find the weights and abscissas given the moments, the nonlinear algebraic equation system shown in \eqref{eq:QuadratureApproximationMoments} must be solved. The combination of \eqref{eq:QuadratureApproximationMoments} with the differential equations of \eqref{eq:SMOMODEs} leads to a differential algebraic equation system which can be solved directly using solvers for differential-algebraic systems (DAE) \cite{gimbunSimultaneousQuadratureMethod2009}. An alternative is the product-difference algorithm (PD) \citep{gordonErrorBoundsEquilibrium1968} as shown in \cite{mcgrawDescriptionAerosolDynamics1997}, which solves \eqref{eq:QuadratureApproximationMoments} efficiently given the moments. Using the approximation of the number distribution function shown in \eqref{eq:QuadratureApproximationDistribution}, the integrals needed for agglomeration in \eqref{eq:PBESourceTerms} can be transformed into sums. The calculation of agglomeration using the quadrature method of moments can be found in \cite{marchisioQuadratureMethodMoments2003, marchisioQuadratureMethodMoments2003a}.

The quadrature method of moments results in a closed equation system for the general PBE from \eqref{eq:PBEGeneral}. In contrast to the standard method of moments, however, it is only an approximation. In addition, to setup the differential equations as in \eqref{eq:SMOMODEs}, at each time step the PD-algorithm needs to be performed, leading to a slightly higher complexity compared to the standard method of moments.

\subsection{Discretized population balance equation (DPBE)}\label{subsection:DPBE}

For applications where knowledge of the full number density function is desired, one has to resort to other methods. The discretization of the distribution is a method that yields the evolution of the full distribution. The most popular methods here are finite volume methods. These methods discretize the domain of $L$ into $N_D$ finite volumes or cells of width $\Delta L _i$ where the average value of the number density in cell $i$ is given by:
\begin{equation}\label{eq:DiscretizedNumberDensity}
    n_i(t)=\frac{1}{\Delta L_i}\int_{L_i^-}^{L_i^+} n(L,t)dL,
\end{equation}
where $L_i^+$ and $L_i^-$ correspond to the faces of cell $i$, i.e., the particle length at the boundaries of cell $i$. Thus, the method leads to a system of $N_D$ ordinary differential equations. Exemplary application to the PBE for cell $i$ considering only growth leads to:
\begin{equation}\label{eq:DiscretizedNumberDensityFunction}
    \frac{\partial n_i(t)}{\partial t} + \frac{G(L_i^+)n(L_i^+,t)-G(L_i^-)n(L_i^-,t)}{\Delta L_i}=0 .
\end{equation}
The accuracy of the method is highly dependent on the method used to determine the values of $n(L_i^+,t)$ and $n(L_i^-,t)$ on the cell faces. Higher-order approximations offer greater accuracy but can lead to oscillations. As shown by \citet{liuWeightedEssentiallyNonoscillatory1994}, a weighted essentially non-oscillatory (WENO) scheme can be used to calculate the fluxes between finite volumes using high orders but still avoiding oscillations. Popular alternatives that are commonly used for PBEs are high-resolution finite volume methods \cite{qamarComparativeStudyHigh2006}. Here, the fluxes are also computed using higher order approximations, but in contrast to WENO, the non-oscillatory property is achieved by using flux-limiters. In the finite volume method, nucleation is easily considered by adding the respective source term scaled by the width of the first cell to the differential equation of the first cell. Agglomeration is considered by computing the birth and death terms of \eqref{eq:PBESourceTermssubeq:Birth} and \eqref{eq:PBESourceTermssubeq:Death} for each element. Performing this calculation for each cell is expensive.

An advantage of the discretization method is the possibility of using a geometric grid. Geometric grids are especially advantageous when particle sizes vary over multiple orders of magnitudes. In addition, they coincide with the nature of agglomeration, where two particles of the same mass form one particle of twice the mass. The calculation of agglomeration for a geometric grid in which the volume is doubled for each succeeding cell is shown in \cite{hounslowDiscretizedPopulationBalance1988}. For some applications, it might be interesting to use a finer grid. Building on the work of Hounslow in \cite{hounslowDiscretizedPopulationBalance1988}, \citet{litsterAdjustableDiscretizedPopulation1995} developed a scheme for a geometric grid where the volume of a succeeding cell is $2^{1/q}$ times the volume of the preceding cell, where $q$ is an integer value.

\subsection{Orthogonal collocation on finite elements (OCFE)}\label{subsection:OCFE}

Another method that yields the full distribution function $n(L,t)$ is orthogonal collocation on finite elements. The domain of the inner coordinate is discretized into finite elements, and the distribution function within each element is approximated by a polynomial. \citet{gelbardNumericalSolutionDynamic1978} first proposed this methodology for PBEs using cubic polynomials. We show the application using Lagrange polynomials for the approximation as described in \cite{alexopoulosPartDynamicEvolution2004} and \cite{alexopoulosPartIIDynamic2005}. An approximate solution of $n(L,t)$ can be obtained by enforcing the original PBE at some collocation points within each finite element. Additionally, algebraic states can be introduced to enforce continuity and differentiability between finite elements. The introduction of algebraic states leads to the necessity of boundary equations. For nucleation-growth processes, the boundary equation for the first element can be easily found as $n(L=0,t)=\frac{N}{G}$. In general, the boundary equation must be chosen according to the process. Some suitable boundary equations can be found in the literature \cite{alexopoulosPartDynamicEvolution2004}.

Similarly to the discretization method, it is possible to define a geometric grid for the finite elements. Unfortunately, solutions for a geometric grid were found to be prone to instabilities. Additionally, by introducing algebraic equations, the OCFE method leads to a DAE system. The necessity of using a usually slower solver for DAE systems is a disadvantage of the method.

\subsection{Monte Carlo (MC)}\label{subsection:MCSim}

Monte Carlo simulations represent an entirely different approach compared to previous solution methods. Instead of tracking the distribution function $n(L,t)$, a large number of individual particles is generated and nucleation, growth, and agglomeration are simulated for the population of particles.

\blue{Methods for Monte Carlo simulations are generally classified into time-driven and event-driven approaches, which differ fundamentally in how they handle time. An overview can be found in \cite{zhaoAnalysisFourMonte2007}. In time-driven methods, a fixed simulation time step is chosen prior to the simulation. At each time step, the extent of nucleation, growth for each particle, and agglomeration during the previous time step is calculated and the population is updated accordingly. In event-driven simulations, time advances based on the occurrence of individual crystallization events. For example, an event could be the agglomeration of two particles. Starting from an initial population, the simulation advances to the first event, processes it (e.g. selecting two random particles for agglomeration and combining them) and advances the simulation time to when that respective event occurs. The simulation then proceeds from one event to the next. Both approaches offer distinct advantages. Event-driven methods are computationally efficient when events are rare as they avoid unnecessary calculations during inactive periods. They also process each discrete event at its exact occurrence time, eliminating discretization errors for stochastic processes. However, time-driven methods are often more suitable for control applications, where the system state is needed at regular, predetermined intervals that naturally align with measurement sampling and controller update cycles.}

\blue{A practical challenge which arises when using Monte Carlo methods for PBEs is to keep track of the total number of simulated particles. As nucleation leads to a population growing in size and agglomeration leads to the population shrinking, it is numerically possible to obtain an excessively large population or a population which is too small. Excessively large populations become computationally expensive, whereas small populations lead to a high statistical variance, compromising the validity of the simulation. Constant-N methods offer a solution to address this challenge by fixing the population to a specific size and assigning weights to the particles of the population \cite{smithConstantnumberMonteCarlo1998}. Subsequently, in the simulation, only the weights are changed with respect to the crystallization phenomena, keeping the size of the population constant through the simulation.}

\blue{Since Monte Carlo methods are very broad and simple in their formulation, they can be applied easily to all various PBE formulations, for example also multi-dimensional PBEs. However, the computational complexity increases significantly with more dimensions as it suffers from the curse of dimensionality, requiring exponentially more particles to maintain statistical accuracy when tracking properties such as size, shape, and composition simultaneously.}

For this work, we present the use of a simple time-driven simulation technique proposed in \cite{vanpeborghgoochMonteCarloSimulation1996}. Here, a time step $\Delta t$ is chosen before the simulation and the manipulation of the particles is performed at each time step. Growth is considered using an Euler method formula:
\begin{equation}\label{eq:GrowthinMC}
    L_{i,t+1}=L_{i,t}+G(L_{i,t}) \Delta t ,
\end{equation}
where $L_i$ represents the length of the $i$-th particle of the distribution. For the case of size-independent growth, this formulation is exact.

For nucleation and agglomeration, particles must be added or deleted from the simulated population of particles. Given the nucleation rate and the agglomeration rate, it is possible to determine the necessary number of nucleation events $N_{\text{nucl}}$ and agglomeration events $N_{\text{agg}}$ for a time step $\Delta t$. Then, for nucleation, $N_{\text{nucl}}$ particles at $L = 0$ are added to the population. For agglomeration, we consider a constant kernel function. Hence, $N_{\text{agg}}$ times, two random particles of the population are chosen, removed from the population, and a single new particle is added to the population. The internal coordinate $L$ of the new particle must be calculated in accordance with the laws of mass conservation.

In the scope of this work, we use the Monte Carlo method to verify the other solution methods presented in cases where no analytical solution exists.

\subsection{Comparison of population balance equation solution methods}\label{subsection:ComparisonSolutionMethods}

The presented solution methods are now compared in three different numerical investigations, providing a brief overview of the strengths and weaknesses of each method. The parameters for each case can be seen in Table~\ref{tb:CaseParameters}. The parameters used in the simulations and the number of time steps have been chosen to show the differences between the methods. Interactive plots of all cases can be found in the dashboard. Additionally, for the first two cases exact solutions are available.
\begin{table}
\caption{Parameters used for case studies. The parameters $N$, $G$, and $\beta$ represent the nucleation rate, growth rate, and the agglomeration rate.}\label{tb:CaseParameters}
\begin{tabular*}{\tblwidth}{@{} LLLLLLL@{} }
\toprule
 & G & N & $\beta$ & Kernel & Time & Initial distribution \\
\midrule
Case 1 & 1.0 & 0 & 0 & None & 15 & Exponential \\
Case 2 & 0 & 0 & 0.5 & Constant & 5 & Exponential \\
Case 3 & 1.0 & 0.01 & 0.1 & Constant & 10 & Gaussian \\
\bottomrule
\end{tabular*}
\end{table}
To establish a connection between numerical investigations and real systems, Table~\ref{tb:CaseRealApplications} lists models from the literature that have the same structure as the models presented here for numerical analysis.
\begin{table}
    \centering
    \begin{threeparttable}
        \caption{Exemplary connection between numerical case studies and models from literature used for real crystallization systems.}\label{tb:CaseRealApplications}
        \begin{tabularx}{\tblwidth}{@{} LL@{} }
            \toprule
             & System \\
            \midrule
            Case 1 & Paracetamol/ethanol \cite{mitchellEstimationGrowthKinetics2011}, KDP/water \cite{eisenschmidtFaceSpecificGrowthDissolution2015}\\
            Case 2 & Particle coalescence \cite{lazzariInterplayAggregationCoalescence2015} \\
            Case 3 & L-alanine/water \cite{kufnerModelingContinuousSlug2023}, sodium chloride/water \cite{rohaniNucleationGrowthAggregation1993}\\
            \bottomrule
        \end{tabularx}
    \begin{tablenotes}
        \footnotesize
        \item KDP = potassium dihydrogen phosphate
    \end{tablenotes}
    \end{threeparttable}
\end{table}

For the DPBE and the OCFE methods, the solution domain is truncated, and the minimum and maximum particle size are chosen to be equal for both methods. For the DPBE method, the solution domain is discretized into finite volumes and the $5$-th order WENO scheme is used. Monte Carlo simulations are conducted using $50 \, 000$ particles for each simulation. The moment-based results are produced using the SMOM approach for case 1, and QMOM for cases 2 and 3. Due to the closure problem, the SMOM approach is not applicable in the latter two and for case 1, QMOM achieved identical results as the SMOM method. For moment-based solutions, we indicate the predicted mean value and $\pm 1$ times the standard deviation in the plot. Note that for the moment-based solution, the information of the moments may be used directly, which may be sufficient for some applications.

\begin{figure}
    \centering
    \begin{subfigure}{\linewidth}
        \includegraphics{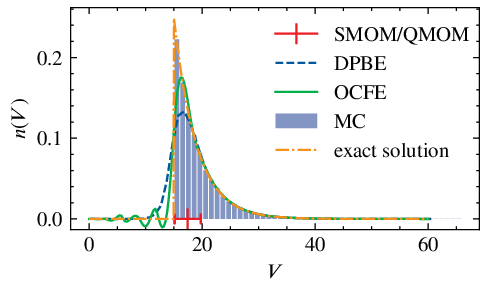}
	   \caption{Case 1: for DPBE and OCFE $60$ differential states were used, where for OCFE, $20$ finite elements with $5$ collocation points each and an artificial diffusion coefficient of $D_a = 0.01$ were used.}
        \label{figure:Case1Distributions}
    \end{subfigure}
    
    \begin{subfigure}{\linewidth}
        \includegraphics{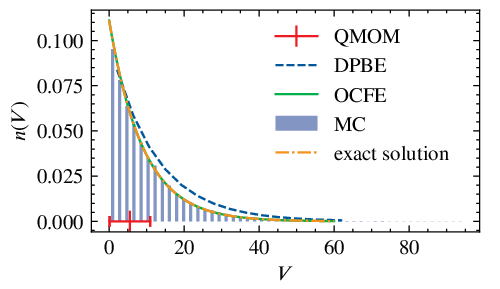}
	   \caption{Case 2: DPBE and OCFE results were obtained using $15$ differential states, with $5$ finite elements á $5$ collocation points for OCFE ($D_a = 0.01$).}
        \label{figure:Case2Distributions}
    \end{subfigure}
    
    \begin{subfigure}{\linewidth}
        \includegraphics{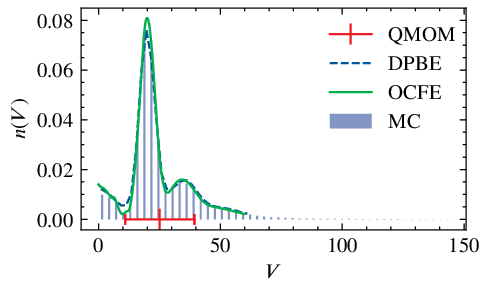}
	   \caption{Case 3: the DPBE and OCFE solution were obtained using $30$ differential states. For OCFE, $10$ finite elements with $5$ collocation points were used ($D_a = 0.01$). Note that the results are normalized for better comparison with the MC solution. Therefore, the boundary condition (N/G) cannot be observed in the figure.}
        \label{figure:Case3Distributions}
    \end{subfigure}
    \caption{Results for the numerical investigation of the different solution methods.}
\end{figure}

The results for case 1 (pure growth) are shown in Figure~\ref{figure:Case1Distributions}.
Numerically, the computation of pure constant growth, given the exponential initial distribution, is very challenging, as it results in a moving discontinuous front. As expected, Monte Carlo methods can model this very accurately, whereas DPBE and OCFE suffer from numerical diffusion. OCFE additionally leads to quite large oscillations, which in turn can be prevented by adding artificial diffusion to the solution. By producing an inherently stabler solution, DPBE is naturally better suited for this problem. However, by choosing a good value for the artificial diffusion parameter, OCFE can lead to stable and accurate results.

The results for case 2 (pure agglomeration) are shown in Figure~\ref{figure:Case2Distributions}. Because the solution is very smooth, DPBE also leads to good results. However, despite only using a few differential states, the accuracy of the OCFE solution is very high. Again, the MC method provides excellent results, and the moment-based solution is not capable of providing meaningful information about the shape of the distribution. However, the moments themselves are accurately captured by the QMOM method.

 The last comparison involves constant nucleation, constant crystal growth, and constant agglomeration. The simulations are conducted for a Gaussian initial distribution. Since no exact solution exists, the MC simulation can be used to compare the accuracy of the other solutions. The results are shown in Figure~\ref{figure:Case3Distributions}. Both DPBE and OCFE lead to accurate results for the case shown in Figure~\ref{figure:Case3Distributions}. Since the solution for this explicit case does not exhibit moving fronts, OCFE is slightly more accurate. However, the bounded domain is a disadvantage of both methods. Because agglomeration leads to results that span multiple orders of magnitude in particle size, the chosen domain quickly becomes too small. OCFE has been found to be unstable when a geometric spacing is used. Hence, DPBE using a geometric grid should be the preferred method for cases like this. The moment-based solution again cannot provide any information about the shape of the distribution.

In summary, moment-based methods are the simplest and provide a good description of the moments. If the full distribution is of interest, other methods must be used. The discretization into finite volumes provides a good compromise between stability and accuracy and is advantageous for most processes. For processes dominated by agglomeration, OCFE leads to high accuracy even with very few states. Monte Carlo simulations are very accurate with a sufficient number of particles examined but are also computationally expensive.

\awesomebox[noteblue]{2pt}{\faLightbulb}{noteblue}{
\textbf{\blue{Takeaways}}
\begin{itemize}[leftmargin=*]
    \item \blue{Choose moment-based methods when only aggregate properties are needed}
    \item \blue{Use the discretized population balance equation (DPBE) with geometric grids for systems with agglomeration}
\end{itemize}
}

\section{Models for different crystallization concepts}\label{section:Models}

Crystallization design targets specific particle characteristics, using the PBE to predict and control these properties. However, the parameters that characterize phenomena such as nucleation, growth, and agglomeration are functions of the state of the continuous phase. Hence, the states that are necessary to model the significant phenomena must be identified and modeled. For cooling crystallization, for example, the necessary states are temperature and concentration as well as potential additional parameters such as stirrer speed or flow velocity. Generally, the methods presented in Section~\ref{section:SolutionMethods} can be coupled with arbitrary models that describe the continuous phase. In this chapter, we show how models for three different crystallizer concepts can be setup and solved: a mixed suspension, mixed product removal (MSMPR) crystallizer, an MSMPR cascade, and a tubular crystallizer. For both MSMPR models we use the assumption that particles are uniformly distributed in space, which allows the usage of the PBE from \eqref{eq:PBEGeneral}. For the tubular crystallizer, we use a finite-volume discretization scheme to solve the partial differential equation, where particles are assumed to be uniformly distributed within each volume, again allowing the usage of \eqref{eq:PBEGeneral}. For all models, we assume the feed flow to contain seed crystals.
\begin{figure}
\centering
\begin{subfigure}{.5\linewidth}
    \centering
    \includegraphics[width=.9\linewidth]{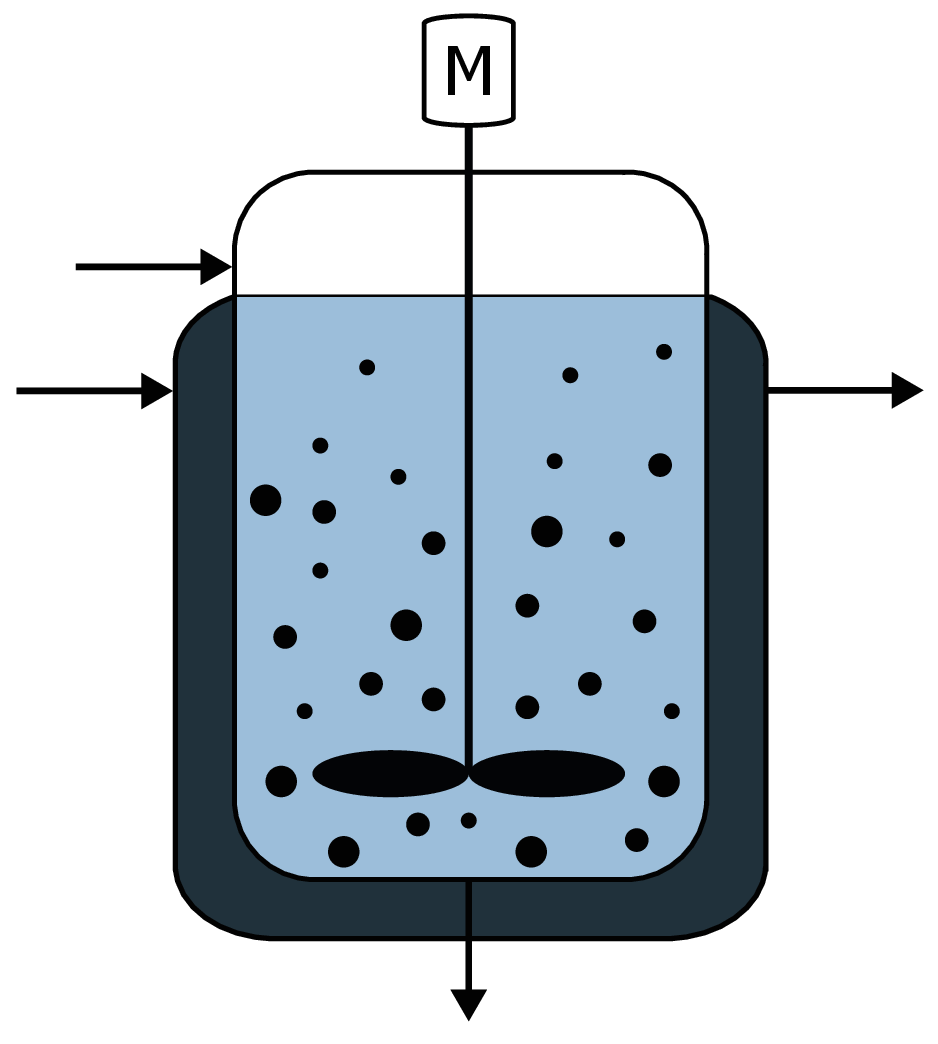}
    \caption{Single-stage MSMPR.}
    \label{figure:MSMPR}
\end{subfigure}%
\begin{subfigure}{.5\linewidth}
    \centering
    \includegraphics[width=.9\linewidth]{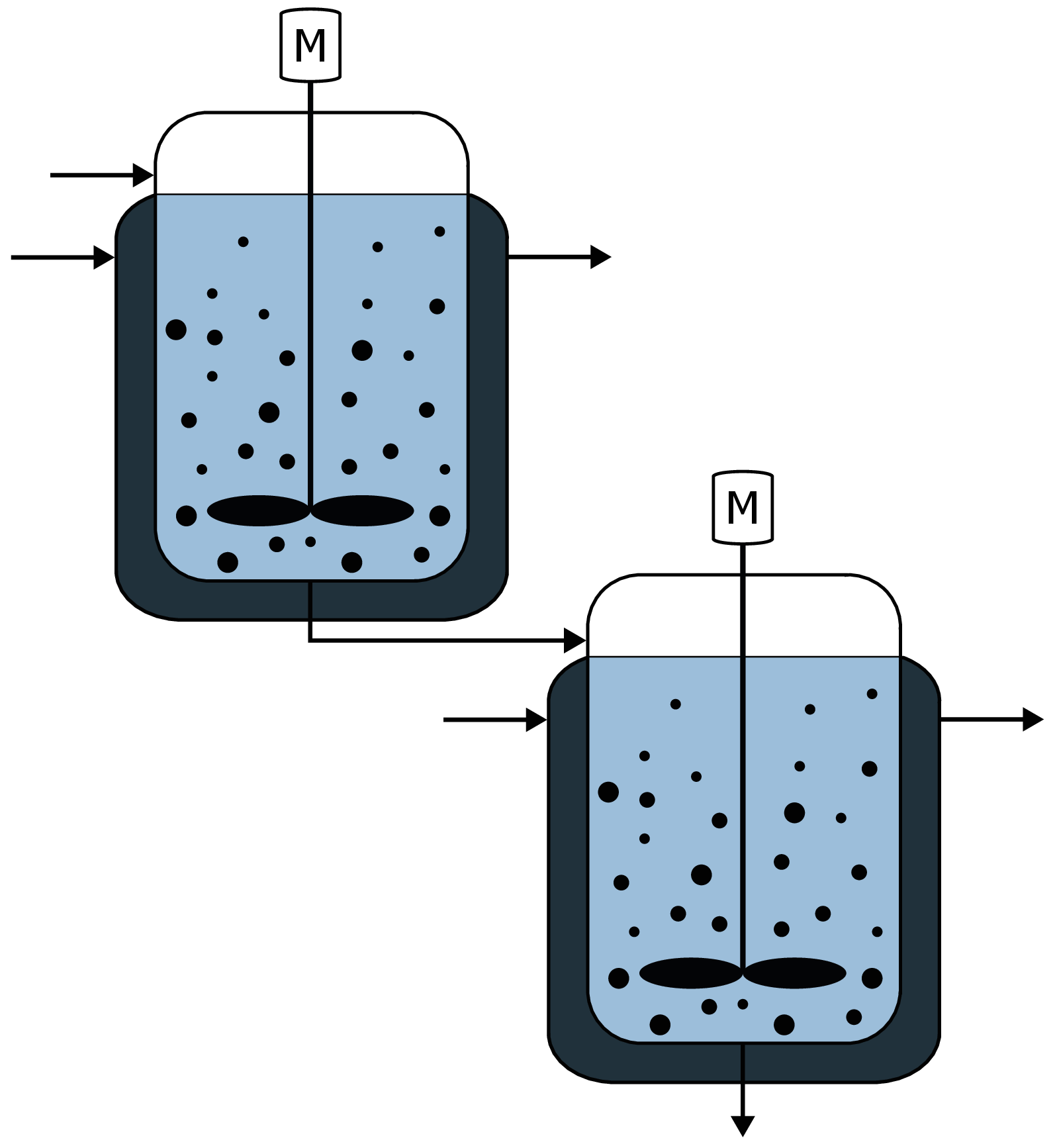}
    \caption{Two-stage MSMPR.}
    \label{figure:multistageMSMPR}
\end{subfigure}

\vspace{1em}

\begin{subfigure}{\linewidth}
    \centering
    \includegraphics[width=\linewidth]{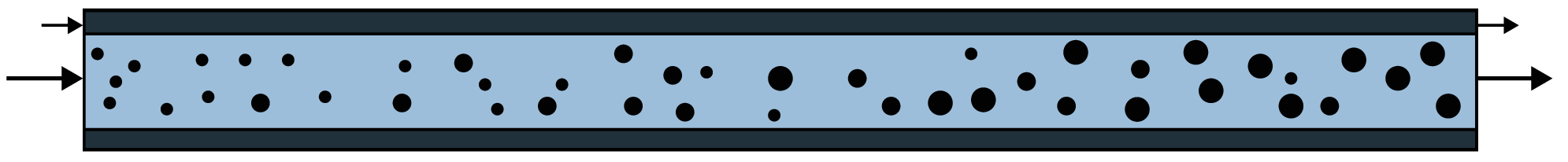}
    \caption{Tubular crystallizer.}
    \label{figure:tubularCrystallizer}
\end{subfigure}

\caption{Sketches of different continuous crystallizer concepts.}
\label{figure:crystallizers}
\end{figure}

\subsection{Mixed-suspension, mixed-product-removal crystallizer}

The MSMPR crystallizer presented here consists of a continuous stirred tank and a cooling jacket, where crystallization occurs only in the stirred tank. Both are assumed to be ideally mixed. A sketch of the MSMPR can be seen in Figure~\ref{figure:MSMPR}. The model equations are setup using a component balance for the concentration $c$ within the crystallizer and energy balances for the temperature within the crystallizer $T$ and within the cooling jacket $T_J$, together with the PBE:
\begin{subequations}\label{eq:MSMPRModelEquations}
\begin{align}
    \frac{d c}{d t} &= \frac{1}{m}(- \dot m_{\text{cryst}}+\rho  F_{\text{feed}}(c_{\text{feed}}-c)), \\
    \frac{d T}{d t} &= \frac{1}{m \, c_p}(- \Delta H_{\text{cryst}} \dot m_{\text{cryst}} \\ \nonumber
    & \quad + \rho  F_{\text{feed}} c_p (T_{\text{feed}}-T) -UA(T-T_J)), \\ 
    \frac{d T_J}{d t} &= \frac{1}{m_J \, c_{p,J}}(\rho_J F_J c_p (T_{J,\text{in}}-T_J) \\ \nonumber
    & \quad -UA(T_J-T)), \\
    \frac{\partial n}{\partial t} &= - G\frac{\partial n}{\partial L} + N + \frac{\rho  F_{\text{feed}}}{m}(n_{\text{initial}}-n),
\end{align}
\end{subequations}
where the properties of the feed are denoted by the subscript $\text{feed}$ and consist of the input flow $F_{\text{feed}}$, the concentration $c_{\text{feed}}$, and the temperature $T_{\text{feed}}$. The particle size distribution is represented by $n$. The properties of the fluid within the crystallizer are the mass $m$, the density $\rho$, and the specific heat capacity $c_p$. Similarly, $m_J$, $\rho_J$, and $c_{p,J}$ represent the mass, the density, and the specific heat capacity of the cooling medium. The mass flow from the continuous phase to the particulate phase is given by $\dot m_{\text{cryst}}$ and the heat of crystallization is given by $\Delta H _{\text{cryst}}$. Finally, the energy transfer between the crystallizer and the cooling jacket is calculated using the heat transfer coefficient $U$ and the area of heat transfer $A$. To keep the notation clear, we show the case of constant $\rho$ and $c_p$. The model can be adapted by using mixing rules to determine the density and specific heat capacity.

A sketch of the two-stage MSMPR can be seen in Figure~\ref{figure:multistageMSMPR}. Each individual MSMPR is modeled as in \eqref{eq:MSMPRModelEquations}. The respective individual MSMPRs are connected by the feed properties. For the first MSMPR the feed properties are equal to \eqref{eq:MSMPRModelEquations}, for the subsequent MSMPRs, the feed temperature and concentration are given by the respective state of the previous MSMPR.

\subsection{Tubular crystallizer}

The tubular crystallizer we present consists of a tube-in-tube design as seen in \cite{termuhlenContinuousSlugFlow2021}. A sketch can be seen in Figure~\ref{figure:tubularCrystallizer}. We assume the flow in the outer tube, containing the cooling medium, to flow co-currently to the flow within the inner tube. Nevertheless, the model of the cooling tube can easily be adapted to model counter-current flow, or even consisting of multiple segments running co-currently or counter-currently. The model of the tubular crystallizer is represented by the convection-diffusion-reaction equation, where crystallization can be interpreted as the reaction:
\begin{subequations}\label{eq:TubularModelEquations}
\begin{align}
    \frac{\partial u}{\partial t} &= -v \frac{\partial u}{\partial z}+D \frac{\partial ^2 u}{\partial z ^2} - s(u), \\
    \frac{\partial n}{\partial t} &= -G\frac{\partial n}{\partial L}-v\frac{\partial n}{\partial z} + D\frac{\partial^2 n}{\partial z^2}+N,
\end{align}
\end{subequations}
where $u$ can represent the concentration or temperature in the process medium and the temperature in the cooling medium. The particle size distribution is represented by $n$. The respective sink and source terms $s(u)$ contain the mass flow from the continuous phase to the discrete phase for the component balance and the heat flows for the heat balances. The flow velocity $v$ in the convection term and the diffusion coefficient $D$ in the diffusion term are considered to be constant along the spatial coordinate $z$. 

\subsection{Practical model implementation and chemical system for case studies}\label{section:ModelImplementation}

\blue{In the following subsections regarding surrogate modeling and MPC, three different crystallization systems are used to demonstrate the presented workflows. The three different example systems are chosen to differ in degree of nonlinearity and complexity, i.e. number of states, to showcase best the advantages and disadvantages of the following methods. For all crystallization systems, the chemical system L-alanine and water is used. The functions and correlations for the crystallization kinetics are shown in Table~\ref{tab:correlations}. Note that agglomeration depends on supersaturation, but not on the internal particle length. Therefore, the kernel is denoted as constant.}
\begin{table}
\centering
\caption{Model correlations for the system L-alanine and water.}
\begin{tabular}{@{}ll@{}}
\toprule
Correlation & Expression \\
\midrule
Growth rate \cite{hohmannAnalysisCrystalSize2018} & $G=5.857 \times 10^{-5} \Delta S^2 \text{tanh} \left(\frac{0.913}{\Delta S} \right) $ \\
Solubility \cite{wohlgemuthInducedNucleationProcesses2012} & $c^*=0.11238 e^{9.0849 \times 10^{-3} T}$ \\
Supersaturation & $\Delta S=\frac{c-c^*}{c^*}$ \\
Beta & $10^{-8}  \Delta S ^ 2  e^{\Delta S}$ \\
Agg. kernel & Constant \\
\bottomrule
\end{tabular}
\label{tab:correlations}
\end{table}

\blue{The first and simplest system is a single-stage MSMPR with the standard method of moments as the solution method for the PBE. In addition, for the model of this system, the functions for solubility and growth rate are linearized. Subsequently, the system is still nonlinear, but the degree of nonlinearity and complexity is the lowest out of the presented systems. The next system is a two-stage MSMPR using the nonlinear solubility and growth rate functions from Table~\ref{tab:correlations}. To keep the number of states low, again the standard method of moments is used for this model. Finally, the last system used to show results is a tubular crystallizer where also the functions from Table~\ref{tab:correlations} are used. In addition, the PBE is solved using the discretized PBE method. First-order approximations are used to solve the convective flow in space and crystal growth in the particle size domain. The resulting model is the most complex of the presented models.}

\awesomebox[noteblue]{2pt}{\faLightbulb}{noteblue}{
\textbf{\blue{Takeaways}}
\begin{itemize}[leftmargin=*]
    \item \blue{Couple continuous phase with particulate phase for full model}
    \item \blue{Model complexity scales with both the finer discretization of space and of the population balance equation (PBE)}
\end{itemize}
}

\section{Transformation to a control-oriented model}\label{section:ControlOriented}

In the previous sections, we showed methods to model the disperse phase, i.e. the crystals, as well as the continuous phase. When solved simultaneously, the full model is obtained. Both modeling components require decisions that balance model accuracy with computational cost. For the dispersed phase, this trade-off involves choosing solution methods and their parameters, such as the number of elements in orthogonal collocation on finite elements. The continuous phase offers fewer modeling choices, though the tubular model still requires selecting the number of finite volumes for discretization.

Detailed models can serve as a digital twin of existing systems, providing real-time insights for process monitoring and controller tuning. In addition, these digital representations facilitate the deployment of advanced control methods. Generally, the model must provide a certain level of accuracy for it to be useful. Secondly, the real-time usage of the model in our control scheme must be feasible. For MPC e.g. the model needs to be suitable for real-time optimization. For some cases, the required level of detail of the model leads to a model that is too complex to be used for real-time control. For these models, the model can be approximated using a data-based surrogate model to keep the required accuracy but achieve real-time feasibility. The price of using an approximation of the model is the limited validity. The data-based model should be used for interpolation only.

\subsection{Surrogate model development}\label{subsection:SurrogateModelDevelopment}

The general structure of approximating the high-fidelity first-principle model that we have obtained from the previous sections is simple. The first-principle model takes current states $x_k$ and inputs $u_k$ and computes the next state $x_{k+1}$:
\begin{equation}\label{eq:InputOutputEquation}
    x_{k+1} = f(x_k,u_k).
\end{equation}
For a single stage MSMPR using a moment-based solution method, the inputs at time step $k$, for example, could be $u_k=\{F_{\text{feed},k}, T_{\text{feed},k}, T_{J,\text{in},k}\}$ and the states at time step $k$ are $x_k=\{c_k, T_k, T_{J,k}, \mu_{i,k}\}$. By gathering data in open-loop simulations, we can subsequently train a data-based model, which approximates $f$ in Equation~\eqref{eq:InputOutputEquation}.

\blue{We present three different methods to obtain a surrogate model. As a linear example, we show how to obtain a surrogate model using subspace identification which leads to a linear surrogate model. As nonlinear methods, we present surrogate neural networks using a nonlinear autoregressive model with exogenous input (NARX) as well as recurrent neural networks (RNN). All methods use past data for their predictions, making the use of measurement data possible.}

\subsubsection{Subspace identification}

\blue{Subspace identification is a data-based method that finds a linear state-space representation given only trajectories of the measurements $y$ and inputs $u$. The goal is to compute the state matrices $A$, $B$, $C$, and $D$ of the following linear state-space representation:}
\begin{subequations}\label{eq:LinearEquationSID}
\begin{align}
    x_{k+1} = Ax_k + Bu_k, \\
    y_{k} = Cx_k + Du_k.
\end{align}
\end{subequations}
\blue{Consequently, given training data consisting of measurements $y$ and inputs $u$, a subspace of state $x$ is computed. The measurement and input data are arranged in block Hankel matrices. Exemplary, for $u_k \in \mathbb{R}^m$, $y_k \in \mathbb{R}^l$, $k=1,\ldots,N+s-1$ the block Hankel matrices for $s=3$ (3 rows in the block Hankel matrix) are given by:}
\begin{align}\label{eq:BlockHankelU}
    U_H &= \begin{pmatrix}
        u_0 & u_1 & u_2 & \cdots & u_{N-1} \\
        u_1 & u_2 & u_3 & \cdots & u_{N} \\
        u_2 & u_3 & u_4 & \cdots & u_{N+1}
    \end{pmatrix},
\end{align}
\begin{align}\label{eq:BlockHankelY}
    Y_H &= \begin{pmatrix}
        y_0 & y_1 & y_2 & \cdots & y_{N-1} \\
        y_1 & y_2 & y_3 & \cdots & y_{N} \\
        y_2 & y_3 & y_4 & \cdots & y_{N+1}
    \end{pmatrix}.
\end{align}
\blue{Given the block Hankel matrices the data equation connecting measurements, states, and inputs can be setup as:}
\begin{align}\label{eq:DataEquation}
    Y_H &= \mathcal{O}X_H+\mathcal{T}U_H \\
\text{with} & \, \, \, \mathcal{O}=\begin{pmatrix}
    C \\
    CA\\
    CA^2
\end{pmatrix}, \mathcal{T}=\begin{pmatrix}
    D & 0 & 0 \\
    CB & D & 0 \\
    CAB & CB &D
\end{pmatrix}, \nonumber \\
    X_H &= \begin{pmatrix}
        x_0 &
        x_1 &
        x_2 &
        \cdots &
        x_{N-1}
    \end{pmatrix}. \nonumber
\end{align}
\blue{where $\mathcal{O}$ is referred to as the extended observability matrix. It is important to note that the identified state space coordinates given by \eqref{eq:LinearEquationSID} are not unique and may differ from physical states by a similarity transformation $T$. The identified matrices ($A_T$, $B_T$, $C_T$, and $D_T$) represent the system in this transformed coordinate system. There exist several methods to extract the system matrices numerically efficient. Popular algorithms are MOESP and N4SID \cite{verhaegenIdentificationDeterministicPart1994, vanoverscheeN4SIDSubspaceAlgorithms1994}. The simplest way to obtain the system matrices is to first eliminate the influence of the inputs by a projection matrix $\Pi^\perp$:}
\begin{align}\label{eq:ProjectionMatrix}
    \Pi^\perp &= I-U_H^T(U_HU_H^T)^{-1}U_H.
\end{align}
\blue{Multiplying the data equation from \eqref{eq:DataEquation} by $\Pi^\perp$ leads to:}
\begin{align}
    Y_H\Pi^\perp &= \mathcal{O}X_H\Pi^\perp. \label{eq:InputsRemoved}
\end{align}
\blue{Subsequently, the column range of $\mathcal{O}$ can be determined by applying the singular value decomposition (SVD) on \eqref{eq:InputsRemoved}. After choosing the desired number of dimensions of the subspace, i.e. the number of states, it is possible to reconstruct the system matrices $A_T$ and $C_T$:}
\begin{align}
    U, S, V_t &= \text{SVD}(Y_H\Pi^\perp), \label{eq:SVD} \\
    U &= \begin{pmatrix}
        C_T \\
        C_T A_T \\
        C_T A_T^2
    \end{pmatrix}. \label{eq:UfromSVD}
\end{align}
\blue{Then, $C_T$ can be extracted directly from $U$, and $A_T$ can be found by solving the least squares problem connecting the first and second row of $U$. The system matrices $B_T$ and $D_T$, and $x_0^{\text{train}}$ (first state of the training data) can be calculated using the data equation from \eqref{eq:DataEquation}. By solving the following system of linear equations using least squares, $B_T$ and $D_T$ can be obtained:}
\begin{subequations}\label{eq:LinearEqSystemSubID}
    \begin{align}
        y_0 &= C_Tx_0^{\text{train}}+D_Tu_0 \\
        y_1 &= C_T(A_Tx_0^{\text{train}} + B_Tu_0)+D_Tu_1 \\
        y_2 &= C_T(A_T^2x_0^{\text{train}}+A_TB_Tu_0+B_Tu_1)+D_Tu_2 \\
         &\quad \quad \quad \quad \quad \quad \quad \, \vdots \nonumber \\
        y_k &= C_T(A_T^kx_0^{\text{train}}+\sum_{\tau=0}^{k-1} A_T^{k-\tau-1} B_T u_\tau)+D_Tu_k.
    \end{align}    
\end{subequations}
\blue{To make predictions starting from an initial measurement $y_0$, an initial state $x_0$ must be calculated. There are several methods that can be used in practice. Easy practical methods are the initialization with zeros or using the solution of the least-squares problem $x_0 \approx C^\dagger(y_0-Du_0)$, where the superscript $^\dagger$ represents the pseudo-inverse. In addition, an observer can be designed to ensure convergence to the true state.}

\awesomebox[nutshellgray]{2pt}{\faCogs}{nutshellgray}{
\textbf{Subspace identification in a nutshell}
\begin{enumerate}[label=(\roman*), leftmargin=20pt, align=left, labelsep=0.1em]
    \item Collect input output data
    \item Structure data as in \eqref{eq:BlockHankelU} \& \eqref{eq:BlockHankelY}
    \item Compute projection matrix from \eqref{eq:ProjectionMatrix}
    \item Perform SVD as in \eqref{eq:SVD}
    \item Get $A_T$ and $C_T$ from $U$ \eqref{eq:UfromSVD}
    \item Solve linear equation system for $B_T$ and $D_T$ \eqref{eq:LinearEqSystemSubID}
\end{enumerate}
}

\subsubsection{Neural network-based surrogate models}

\blue{Most systems and first-principle models for crystallization are not linear. For these cases, it is clear that the surrogate model must also be nonlinear to accurately capture the dynamics. First, we present the general concept of neural networks. Then, we show how to arrange the training data for neural network training. Subsequently, we show two different concepts based on neural networks to obtain surrogate models. Nonlinear autoregressive models with exogenous inputs (NARX) \citep{chenNonlinearSystemIdentification1990, billingsNonlinearSystemIdentification2013} based on neural networks and RNNs \cite{rumelhartLearningRepresentationsBackpropagating1986}.}

\begin{figure}
    \centering
    \includegraphics{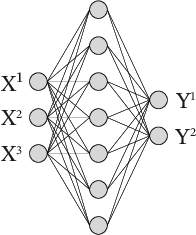}
    \caption{\blue{Sketch of a standard feedforward neural network with one hidden layer. The network consists of $3$ inputs and $2$ outputs. The entries of the input $X$ are denoted as $X^i$ and the entries of the output $Y$ are given by $Y^i$}}
    \label{figure:NeuralNetwork}
\end{figure}

\blue{Figure~\ref{figure:NeuralNetwork} schematically visualizes a neural network consisting of an input layer, a hidden layer, and an output layer.}
\blue{The circles represent neurons. A connection between neurons indicates a flow of information. A neural network approximates the connection between the inputs (here: $X=\left(X^1 \,\, X^2 \,\, X^3\right)$) and the outputs (here: $Y=\left(Y^1 \,\, Y^2\right)$). The inputs are fed on the left in Figure~\ref{figure:NeuralNetwork} and their values are passed to the hidden layer (here consisting of $7$ neurons) where they go through a linear and a nonlinear transformation at each neuron:}
\begin{align}
    a_1^1=g(w_1^1X^1+w_2^1X^2+w_3^1X^3+w_4^1),
\end{align}
\blue{where $w_{1}^1$, $w_{2}^1$, and $w_{3}^1$ represent the weights and $w_4^1$ the bias of the first neuron. The function $g$ is the nonlinear activation function. Typically, $\text{tanh}$, $\text{sigmoid}$, or $\text{ReLU}$ are used for $g$. Here, $a_1^1$ is the activation of the first neuron on the first layer. Stacking weights and biases as well as the activations in matrices and vectors lead to the following function composition for the full network:}
\begin{subequations}
\begin{align}\label{eq:NN}
    a_{1}&=g(W_{0}X), \\
    Y&=W_{1}a_1,
\end{align}
\end{subequations}
\blue{where the variable $a_1$ contains the activations of the neurons of the hidden layer. The weight matrices $W_i$ contain the weights and biases. Typically, for regression, the output layer consists only of a linear transformation. Training the neural network means changing the weights and biases so that the predicted $Y$ is close to the true $Y_{\text{true}}$, which is determined by the objective function. A typical objective function is the mean-squared-error:}
\begin{align}\label{eq:ObjectiveFunction}
    \min_{w} &\ \frac{1}{N}\sum^N_i \lVert Y_{\text{true},i}- Y_i \rVert _2^2.
\end{align}
\blue{Training is usually performed by backpropagation, where the gradients of the objective function are calculated with respect to the weights. Subsequently, the weights are varied so that the fit is improved.}

\blue{Comparison to \eqref{eq:InputOutputEquation} determines how the simulation data needs to be structured. The training data matrices $X$ and $Y$ for $N$ data points to train the surrogate model are constructed by:}
\begin{align}\label{eq:TrainingData}
    X= \begin{pmatrix}
        x_0 &u_0 \\
        x_1 &u_1 \\
        \vdots & \vdots \\
        x_k & u_k \\
        \vdots & \vdots \\
        x_{N-1} &u_{N-1}
    \end{pmatrix}, \, Y= \begin{pmatrix}
        x_1 \\
        x_2 \\
        \vdots \\
        x_{k+1} \\
        \vdots \\
        x_{N}
    \end{pmatrix}.
\end{align}
For more complex models, i.e. models containing a larger number of states, it might not be desirable to use every state as input to the data-based model and to predict every state as output. To avoid an unnecessarily large data-based model, i.e. a large number of inputs and outputs of the neural network, we can choose to only use a subset of the states. This subset we refer to as measurements $y$. Measurements can be selected such that $y$ corresponds to real measurements. In addition, states that are of interest for process control or process monitoring can be added to $y$. This usually includes all states at the output of the crystallizer. As for the spatial dimension, where we only choose states at certain positions as measurements, it is useful not to predict the entire particle size distribution. Methods that provide the entire particle size distribution allow for the calculation of specific characteristic parameters, for example, the mode or median of the distribution, as well as any desired widths. To reduce the size of the data-based model, $y$ should only contain the characteristic parameters of the particle size distribution that are important for the process. Further reduction of the complexity of the neural network can be achieved by applying principal component analysis to the input of the neural network. \blue{When using only a subset of the states for the next prediction, state information important for the prediction of the next time step may be missing. Therefore, in addition to current measurements and inputs, the model must also use the history of measurements and inputs for the predictions.}

\blue{NARX models encode the history of measurements and inputs directly in the input of the data-based model. Instead of using only $y_k$ and $u_k$ for the prediction of $y_{k+1}$, past measurements and inputs are used additionally as input of the model. Compared to \eqref{eq:InputOutputEquation}, the use of NARX results in the following structure for the prediction of our model:}
\begin{equation}\label{eq:InputOutputEquationNARX}
    y_{k+1} = f(y_k,y_{k-1},\ldots,y_{k-l},u_k,u_{k-1},\ldots,u_{k-l}).
\end{equation}
\blue{Past measurements and inputs are considered up to time step $k-l$ (where the lag parameter $l$ is a degree of freedom). The outputs still consist only of $y_{k+1}$. The significantly reduced number of outputs (predicting only $y_{k+1}$ rather than the full state $x_{k+1}$) makes the learning problem usually much simpler, leading to better prediction accuracies.}

\blue{The training data matrices are constructed similar to \eqref{eq:TrainingData} for NARX models. The $k$-th row is given by:}
\begin{subequations}\label{eq:TrainingDataNARX}
\begin{align}
    X_k &= \begin{pmatrix}
        y_k & y_{k-1} \ldots y_{k-l} & u_k & u_{k-1} \ldots u_{k-l}
    \end{pmatrix}, \\
    Y_k &= \begin{pmatrix}
        y_{k+1}
    \end{pmatrix}.
\end{align}
\end{subequations}
\blue{It is important to note that NARX is not a concept specific to neural networks. It is possible to use any nonlinear function for $f$ in \eqref{eq:InputOutputEquationNARX}.}

\awesomebox[nutshellgray]{2pt}{\faCogs}{nutshellgray}{
\textbf{NARX neural networks in a nutshell}
\begin{enumerate}[label=(\roman*), leftmargin=20pt, align=left, labelsep=0.1em]
    \item Collect input output data
    \item Choose lag parameter $l$
    \item Structure data as in \eqref{eq:TrainingDataNARX}
    \item Train neural network on $X$ and $Y$ minimizing \eqref{eq:ObjectiveFunction}
\end{enumerate}
}

\begin{figure*}
    \centering
    \includegraphics{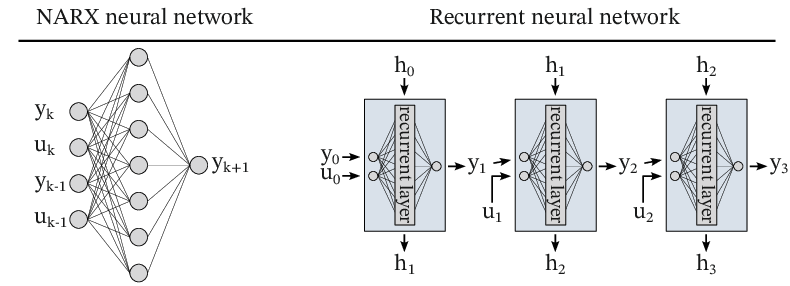}
    \caption{\blue{Visualization of recursive prediction using NARX neural network models and RNN models. A single prediction of the next time step is shown for the NARX neural network. For the RNN, a rollout for three time steps is shown. The same RNN is evaluated three times using updated measurements $y$, inputs $u$, and hidden states $h$.}}
    \label{figure:NARXvsRNN}
\end{figure*}

\blue{In contrast to NARX models where the history of the measurements is encoded directly in the inputs, RNNs represent an alternative. The review in \cite{yuReviewRecurrentNeural2019} provides an overview of RNNs and related more sophisticated approaches. For RNNs, the necessary information about the temporal evolution of the measurements and inputs is encoded in a hidden state. The hidden state is an additional output of the network and is used as an input for the prediction of the next time step. Subsequently, the information about the temporal evolution of measurements and inputs which would be missing is still contained in the hidden state. From an architectural perspective, the RNN adds only a single recurrent weight matrix $W_{hh}$ to the standard feedforward network:}
\begin{subequations}
\begin{align}\label{eq:RNN}
    h_{k+1}&=\text{tanh}(W_{ih,0}X_k+W_{hh,0}h_k), \\
    Y_{k+1}&=W_{ih,1}h_{k+1}.
\end{align}
\end{subequations}
\blue{The activation function is the $\text{tanh}$ function for RNNs. The simple modification of the structure of the network fundamentally changes the capabilities of the network to enable temporal memory. Training uses backpropagation through time, a variant of standard backpropagation where gradients are computed backward through the entire temporal sequence. Training data must be divided into sequences to perform a rollout as seen in Figure~\ref{figure:NARXvsRNN} (right). A single sequence $\text{seq}$ could, for example, be given by:}
\begin{align}\label{eq:Sequence}
    \text{seq} = \begin{pmatrix}
        (X_0, Y_0) & (X_1, Y_1) & (X_2,Y_2) & (X_3, Y_3)
    \end{pmatrix}.
\end{align}
\blue{The length of sequences can be chosen freely. Sequences should not be chosen to short for the model to be able to learn the temporal dependency. However, for long sequences, standard RNNs suffer from vanishing gradients. A rollout consists of a forward pass that predicts the full sequence once. Subsequently, a backward pass is performed, computing the gradient through all time steps of the sequence. The input weights $W_{ih}$ and the recurrent weights $W_{hh}$ are then updated. The input weights learn to extract features from the current input $X_k=(y_k \, \, u_k)$, while the recurrent weights learn to extract the relevant information of $h_k$ for the prediction of $y_{k+1}$.}

\blue{Figure~\ref{figure:NARXvsRNN} schematically represents the difference between NARX neural network models and RNN models for inference.}
\blue{For simplicity, the system consists of a single measurement and a single input. The NARX neural network uses the current measurement $y_k$ and the current input $u_k$ and, in addition, measurement and input of the previous time step $y_{k-1}$ and $u_{k-1}$ as input to predict only the next state $y_{k+1}$. For the next prediction, the predicted $y_{k+1}$, the new input $u_{k+1}$, as well as $y_k$ and $u_k$ are used to predict $y_{k+2}$ and so on. For the RNN, the prediction starts at $y_0$ and $u_0$. Typically, the first hidden state $h_0$ is initialized to zeros and the first evaluation of the RNN yields the next measurement $y_1$ and the next hidden state $h_1$. The RNN is then recursively evaluated using the respective next measurement and the next hidden state as input.}

\awesomebox[nutshellgray]{2pt}{\faCogs}{nutshellgray}{
\textbf{Recurrent neural networks in a nutshell}
\begin{enumerate}[label=(\roman*), leftmargin=20pt, align=left, labelsep=0.1em]
    \item Collect input output data
    \item Structure data as in \eqref{eq:TrainingData}
    \item Divide data into sequences \eqref{eq:Sequence}
    \item Train recurrent neural network on sequences minimizing \eqref{eq:ObjectiveFunction} using backpropagation through time
\end{enumerate}
}
\subsection{Application to example systems}

\begin{figure*}
    \centering
    \includegraphics{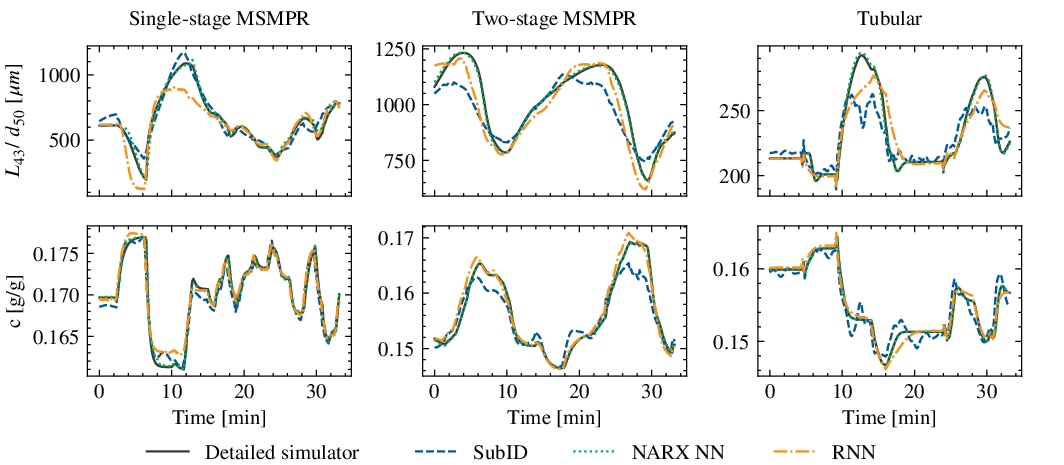}
	\caption{\blue{Comparison of subspace identification (SubID), nonlinear autoregressive models with exogenous inputs using neural networks (NARX NN) and recurrent neural networks (RNN) for surrogate model development on three different systems. The upper row shows the mean diameter $L_{43}$ for the MSMPR models, and the $d_{50}$ for the tubular crystallizer. The models are evaluated as simulation models as shown in \cite{schoukensNonlinearSystemIdentification2019}. The model prediction is recursively fed back as the input for the prediction of the next time step.}}
    \label{figure:ComparisonDifferentModels}
\end{figure*}

\blue{Subsequently, we present the application of the three presented methods for surrogate development to the three example systems of Section~\ref{section:ModelImplementation}. For all NARX models, a lag parameter of $l=10$ is chosen along with a time step of $10 \, s$.} To train data-based models, sufficiently informative training data sets are required. In order to obtain these data sets, random values are sampled from the permissible ranges of the inputs for the models, and these values are applied to the system for different time durations. For all models, the training data set $\{X,Y\}$ contains $100 \, 000$ samples. \blue{Figure~\ref{figure:ComparisonDifferentModels} shows the prediction accuracy of the different surrogate modeling methods applied to the different systems.}
\blue{The average model accuracy of the closed-loop prediction experiments in Figure~\ref{figure:ComparisonDifferentModels} can be seen in Table~\ref{tb:ResultsSurrogateModels}.}
\begin{table}
\caption{Mean squared error (MSE) for the presented surrogate models calculated for the test data shown in Figure~\ref{figure:ComparisonDifferentModels}.}\label{tb:ResultsSurrogateModels}
\begin{tabular}{@{}ll@{}}
\toprule
Method & MSE \\
\midrule
Subspace identification & $0.487$ \\
NARX neural networks & $0.002$ \\
Recurrent neural networks & $0.146$ \\
\bottomrule
\end{tabular}
\end{table}
\blue{NARX neural networks lead to the highest prediction accuracy by a large margin. This superior performance can be attributed to the explicit encoding of temporal dependencies in the input space, which the RNN first has to learn in the hidden state. Nevertheless, RNNs also provide accurate predictions for some parts of the trajectories but exhibit larger errors elsewhere. Linear models obtained by subspace identification lead to good results for the single-stage MSMPR system, which is the least nonlinear. For the other two systems, the results are not satisfactory. For nonlinear systems, the state dimension of the identified linear model needs to be increased to capture the system dynamics, but this often leads to oscillatory or unstable predictions.}

The models were implemented using the tool do-mpc \cite{fiedlerDompcFAIRNonlinear2023} with CasADi \cite{anderssonCasADiSoftwareFramework2019}. The simulation was performed using SUNDIALS solvers \cite{hindmarshSUNDIALSSuiteNonlinear2005}. The neural networks were trained using the PyTorch framework \cite{Ansel_PyTorch_2_Faster_2024}. 

\subsection{Comparative discussion of surrogate model methods}

\blue{Crystallization systems exhibit a high-dimensional state space (distributed in spatial direction and crystal size), they are nonlinear and exhibit time delays (especially for longer cascades and tubular crystallizers). The high-dimensional underlying state space is addressed directly by subspace identification and RNNs, using a potentially higher dimensional state subspace or hidden state, respectively. NARX models encode information about the underlying state space using the history of the measurements, manually constructing a larger state space. Since a nonlinear system can only be accurately approximated by a nonlinear model, subspace identification is only a viable option when the system is linear or close to linear. RNNs and NARX neural networks are suitable for nonlinear systems. RNNs learn to capture the time delay in the hidden state, which should contain the necessary information to predict the next measurements accurately. NARX models directly encode the time delay using the history of measurements and inputs as additional inputs to the model. Since it is not possible to model a time delay directly using a linear model, subspace identification is not suited for systems that exhibit a large time delay. In addition to the mentioned challenges, crystallization dynamics are often stiff, i.e. there are multiple time scales present (crystals grow slower than heat is transferred) making system identification more difficult.}

\blue{With respect to prediction accuracy, NARX neural networks have outperformed subspace identification and RNNs for the investigated systems here. In contrast, linear models as obtained by subspace identification are far simpler to work with, i.e. for the deployment in advanced control or designing an observer. If the underlying system is linear or close to linear and the prediction accuracy using a linear model is sufficient, subspace identification is very promising. If the model accuracy of the linear model does not suffice, from our experience, neural networks using NARX are a well-suited method. It is important to note that more evolved algorithms based on RNNs such as long short-term memory models exist that are more complex but will likely lead to a better prediction accuracy compared to RNNs. An overview can be found in \cite{yuReviewRecurrentNeural2019}. Regardless of the surrogate method chosen, it is important to emphasize that all data-based models are valid only for interpolation within the training data range. Extrapolation beyond the domain of training data can lead to unreliable predictions. This is a disadvantage compared to first-principle models and must be considered in the control algorithm.}

\blue{Beyond the methods presented, there exist other methods for system identification which offer different advantages. Dynamic mode decomposition with control \cite{proctorDynamicModeDecomposition2016} is, for example, a linear alternative to subspace identification, which can also be used for nonlinear systems using Koopman theory \cite{williamsDataDrivenApproximation2015, kordaLinearPredictorsNonlinear2018}. Physics-informed neural networks \cite{raissiPhysicsinformedNeuralNetworks2019} combine the predictive power of neural networks while still enforcing some physical constraints, trading-off accuracy and complexity. In addition, there exists a full spectrum of methods that can quantify the prediction uncertainty. Quantifying prediction uncertainty is beyond the scope of this tutorial overview, but the reader is directed to \citep{schoukensNonlinearSystemIdentification2019} for a full practical overview of system identification.}

\awesomebox[noteblue]{2pt}{\faLightbulb}{noteblue}{
\textbf{\blue{Takeaways}}
\begin{itemize}[leftmargin=*]
    \item \blue{Linear surrogates suffice only for weakly nonlinear systems without time delays}
    \item \blue{Reduce surrogate complexity by predicting only measurements and control-relevant characteristics}
    \item \blue{Surrogate models are only valid for interpolation}
\end{itemize}
}

\section{Model predictive control for crystallization}\label{section:MPC}

The overall objective of this paper is to show the steps of applying MPC to continuous crystallization. The practitioner needs to setup the system model with the objective function of the control scheme in mind. Applications that require control of the shape of the size distribution require solution methods that track the full distribution. Subsequently, it is possible to control some characteristic values of the distribution. In this chapter, we show the application of MPC to the two different models obtained in the previous section. The two-stage MSMPR model was chosen as a low-complexity model, allowing comparison with an MPC controller using the first-principle model. By comparison to this benchmark solution, insights can be gained even though the presented methodology of approximating the model would not be necessary for this case. The second system, the tubular crystallizer using the DPBE method, is an example in which the proposed method is necessary. Due to its complexity, the resulting model is not suitable for optimization in MPC.

\subsection{Model predictive control}

The idea of MPC is to solve an optimization problem that produces an optimal sequence of inputs for a given prediction horizon of $N$ steps and to apply only the first input of the sequence to the system. In the next time step, the optimization problem is solved again with the updated states, and again the first input is applied to the system. This procedure is repeated for the duration of the process. The algorithm therefore inverts the process of simulating a model. Instead of choosing input values and calculating the next states, we choose desired states (or at least a function of the states and inputs) and calculate a sequence of inputs that achieve the desired states in an optimal fashion for the prediction horizon.

The standard optimization problem for nonlinear MPC can be formulated as follows:
\begin{subequations}\label{eq:MPC}
\begin{alignat}{2}
    &\min_{u_k} &\quad &\sum_{k=0}^{N-1} l(s_k, u_k) + V_f(s_N) \\
    &\, \, \, \text{s.t.} & & s_{k+1} = f(s_k, u_k) \\
    &   & &s_k \in \mathcal{S} \\
    &   & &u_k \in \mathcal{U} \\
    &   & &s_0 = s_{\text{initial}},
\end{alignat}
\end{subequations}
where the states of the system at time step $k$ are denoted as $s_k$ and the inputs at time step $k$ are $u_k$. The cost function is divided into stage cost $l$ and terminal cost $V_f$. The system model is given by $f$, state constraints by $\mathcal{S}$, and input constraints by $\mathcal{U}$. Furthermore, a constraint is added for the initial state $s_0$ to be $s_{\text{initial}}$. \blue{We use $s$ for states here to symbolize that the concept of a state here can be applied to the surrogate models shown before, regardless of whether the actual full state of the system $x$ or only measurements $y$ are used.} For the solution of the MPC problem, do-mpc \cite{fiedlerDompcFAIRNonlinear2023} and CasADi \cite{anderssonCasADiSoftwareFramework2019} were used with the IPOPT solver \cite{wachterImplementationInteriorpointFilter2006}.

\subsection{Model predictive control using the first-principle model}

\begin{figure*}
    \centering
    \includegraphics{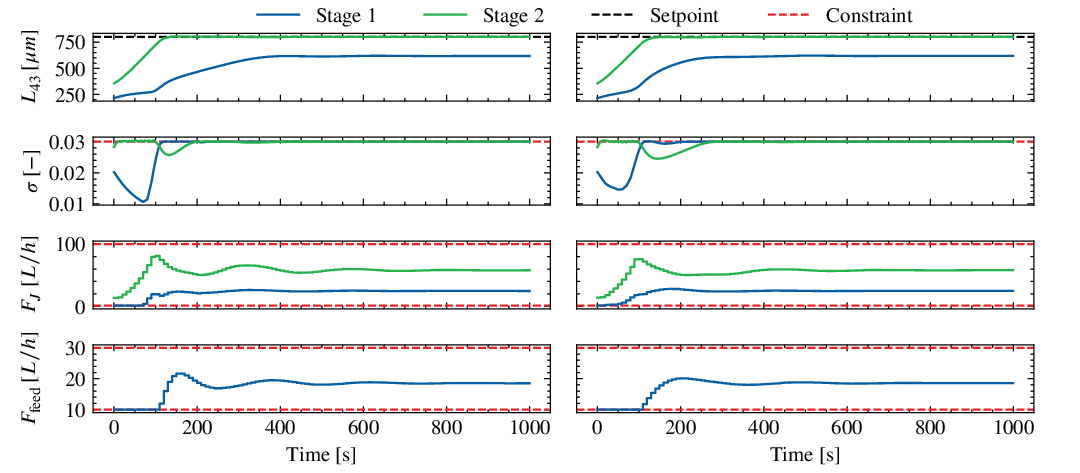}
    \caption{Comparison of the control performance using the first-principle model (left) and the data-based model (right) as internal model for the controller. The system is the two-stage MSMPR, with the blue lines corresponding to the first stage and the green line corresponding to the second stage. The dashed red lines represent input or state constraints. The dashed black line in the plots for $L_{43}$ corresponds to the setpoint.}
    \label{figure:MPCMSMPR}
\end{figure*}

First, we show the results for MPC using the first-principle model. As described before, MPC using the first-principle model is performed only for the two-stage MSMPR model. The goal of the control scheme is to track a certain mean diameter $L_{43,\text{set}}=800 \, \mu m$ at the outlet of the last crystallizer. In addition, the goal is to maximize the yield of the particulate phase. Since we assume the weight fraction of the feed flow to be constant, we maximize the feed flow $F_{\text{feed}}$. As inputs, the three streams to the cooling jackets $F_{J,i}$ and the feed $F_{\text{feed}}$ are used. The cost function is given as:
\begin{subequations}\label{eq:CostFunctionPhysicalModel}
\begin{align}
    l &= w_1 (L_{43}-L_{43,\text{set}}) ^2 - w_2 F_{\text{feed}} \\ \nonumber
        & \quad - w_3 \sum_{i=1}^2 F_{J,i} + \Delta u^T w_4 \Delta u, \\
    V_f &= w_1 (L_{43}-L_{43,\text{set}}) ^2.
\end{align}
\end{subequations}
To obtain a better control performance and achieve a high cooling rate in both stages, we also maximize the feed to the cooling jackets $F_{J,i}$ and introduce a penalty on input changes ($u = (F_{J,1},F_{J,2},F_{\text{feed}})^T$). Finally, to stay within a reasonable region of supersaturation, we impose a soft constraint on the relative supersaturation at $\sigma=0.03$. The results of this case are shown on the left in Figure~\ref{figure:MPCMSMPR}.
Due to the tuning of the cost function in \eqref{eq:CostFunctionPhysicalModel}, the controller first reaches the set mean diameter as quickly as possible. Subsequently, while maintaining the diameter of the particle, the inputs are chosen such that the feed stream $F_{\text{feed}}$ is maximized. The strategy how the model predictive controller achieves this performance becomes apparent. First, the feed flow is kept at the lower bound, as a smaller feed can be cooled more quickly, and the desired diameter can therefore be reached faster. The controller initially cools only in the second stage, precisely enough to work at the supersaturation constraint in the second stage. As soon as the desired diameter is reached at the outlet of the second stage, the controller focuses on maximizing the feed and minimizing the cooling flows. The feed is increased and the cooling flows adjusted so that both stages operate at the supersaturation constraint, which corresponds to optimal operation.

\subsection{Model predictive control using surrogate models}

The results obtained by using the first-principle model within the MPC scheme are now compared to the case where a data-based approximation is used for the model in the MPC controller. For the simulation of the system, the first-principle model is used. The same cost function as before is used \eqref{eq:CostFunctionPhysicalModel}. However, to ensure a fair comparison, the weights of the cost functions were individually tuned to obtain the best results in both cases. The results for the data-based case of the two-stage MSMPR are shown on the right in Figure~\ref{figure:MPCMSMPR}. The controller takes the same approach as before. First, the mean diameter $L_{43}$ is moved to the set value as quickly as possible. Subsequently, the feed is maximized. Approaching the steady state, the optimizer finds the same optimal operating point. Compared to the controller using the first-principle model, the same strategy is followed here.

A common problem when using data-based models in optimization is possible model exploitation by the optimizer. Since the data-based model is only an approximation, the optimizer might find low-cost solutions that are physically unrealistic. These regions can exist, for example, due to limited training data or extrapolation errors. A common solution in the literature is to use a data-based model which can quantify the uncertainty of its predictions and to constrain operation to regions where the data-based model indicates low uncertainty. This can be realized, for example, by constraining the operation to trust regions \cite{fiedlerModelPredictiveControl2022} or using the uncertainty information in multi-stage MPC \cite{johnsonRobustNonlinearModel2024}. In practice, a similar effect can be achieved by penalizing changes of the inputs. Therefore, an additional challenge in adjusting the controller when using a data-based model as an internal model is to adjust the weight $w_4$ from Equation~\eqref{eq:CostFunctionPhysicalModel}. The easiest way for the user to find good values for $w_4$ is to first tune the cost function (that is, select weights $w_1$, $w_2$, and $w_3$) and then increase weight $w_4$ until the input trajectories do not contain overly aggressive changes.

\blue{To present a comprehensive analysis and account for statistical variability when training data-based models, 10 different data-based models were trained on the same data set. Subsequently, the MPC problem using the cost function \eqref{eq:CostFunctionPhysicalModel} is solved using the different models. The results of the simulations are summarized in Table~\ref{tb:ResultsMSMPR}.}
\begin{table}
\caption{\blue{Results for the use of the first-principle model and the data-based neural network models as the internal model in MPC. The use of the first-principle model in MPC, which represents the best possible result in this scenario, is compared to the use of data-based models used in MPC. To account for statistical variability, the performance metrics of the data-based models represent the average value of the 10 different models. The closed-loop cost represents the actual attained objectives (maximizing $F_{\text{feed}}$ and tracking $L_{43}$).}}\label{tb:ResultsMSMPR}
\begin{tabular}{lc@{\hspace{1pc}}cc}
\toprule
 & \makecell[c]{Closed-loop \\ cost} & \makecell[c]{Rel. constraint \\ violation $[\%]$} & \makecell[c]{CPU \\ time [s]} \\
\midrule
First-principle & 79.9 & 0.01  & 0.141  \\
Data-based & 81.1  & 0.07 & 3.054  \\
\bottomrule
\end{tabular}
\end{table}
The results using the data-based models are very close to the benchmark first-principle model MPC solution. \blue{The average closed-loop cost when using the data-based models is only slightly worse than when using the first-principle model, validating the results that can be seen in Figure~\ref{figure:MPCMSMPR}. The relative constraint violation when using the data-based models is slightly higher compared to the first-principle model, but is still at a very low level in general.} Given that the data-based model is computationally more expensive than the first-principle alternative, it does not provide any speed advantage for MPC and should not be used in this application.

\begin{figure}
    \centering
    \includegraphics{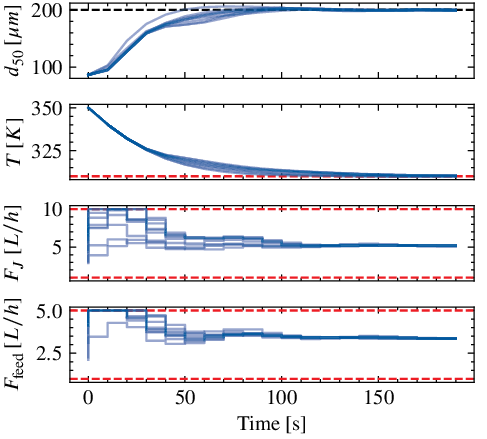}
	\caption{Simulation results of controlling the tubular system using a data-based model in MPC and the first-principle model for the simulator. The results are shown for the tubular crystallizer. The states correspond to the states at the outlet of the crystallizer. The red dotted lines represent input or state constraints. The results are presented using $10$ different data-based models in the controller which were trained on the same data set.}
    \label{figure:MPCTubular10Sims}
\end{figure}

\begin{figure}
    \centering
    \includegraphics{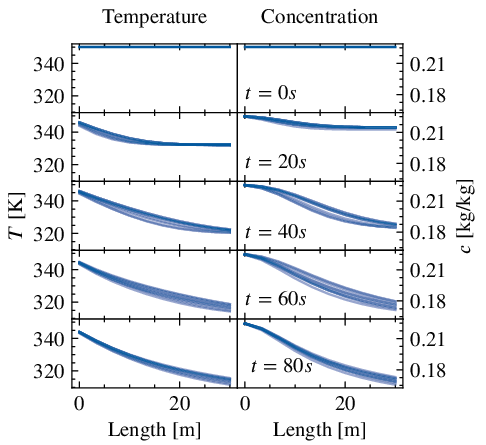}
	\caption{Snapshots of the states of the tubular crystallizer over the spatial length at different times. The left column shows the temperature profiles while the right column represents the concentration profiles.}
    \label{figure:SnapshotsTubular}
\end{figure}

For the case study of the tubular system, the following cost function is used for MPC:
\begin{subequations}\label{eq:CostFunctionSurrogateModelTubular}
\begin{align}
    l &= w_1 (d_{50}-d_{50,\text{set}}) ^2 - w_2 F_{\text{feed}} + \Delta u^T w_4 \Delta u, \\
    V_f &= w_1 (d_{50}-d_{50,\text{set}}) ^2.
\end{align}
\end{subequations}
In contrast to before, we track a particle size for the median diameter $d_{50} = 200 \, \mu m$. The median diameter was not available before because the method of moments was used. In addition, the feed $F_{\text{feed}}$ is maximized. For the temperature, a constraint is imposed at the outlet of the crystallizer at $T=310 \, K$. Again, we penalize changes in the inputs. The results for the tubular case are shown in Figure~\ref{figure:MPCTubular10Sims} and Figure~\ref{figure:SnapshotsTubular}. The results are summarized in Table~\ref{tb:ResultsTubular}. No comparison can be made to the first-principle model, since for the tubular system, the complexity of the first-principle model prevents direct usage in the control algorithm. Therefore, the values for closed-loop cost are difficult to interpret. Fortunately, the controller was able to consistently avoid constraint violations. As for the case study of the MSMPR system, 10 different models are trained on the same data set for the tubular crystallizer. Similarly to the previous system, the controller first attempts to achieve the desired diameter as quickly as possible \blue{and then chooses inputs to operate at the optimal steady state.} The controller can now only select the two flows for the coolant and the feed at the inlet of the crystallizer. The dynamics of the states at the outlet are strongly influenced by the time delay due to the length of the crystallizer. The system is also more complicated to control because the time delay depends directly on the settings of the inputs. This means that a larger cooling flow, for example, cools the system better but also flows at a higher speed, which makes the system faster. The control inputs are therefore not as easy to interpret as before, but the controller still keeps the desired diameter constant and finds an optimal operating point at the temperature constraint.
\begin{table}
\caption{Results using the data-based model as internal model for the MPC controller for the tubular crystallizer system. The complexity of the first-principle model prevents its direct use in the controller for comparison.}\label{tb:ResultsTubular}
\begin{tabular}{@{}ll@{\hspace{1pc}}l@{}}
\toprule
 & \makecell[c]{Rel. constraint \\ violation $[\%]$} & \makecell[c]{CPU \\ time [s]} \\
\midrule
First-principle & \multicolumn{2}{c}{Not solvable} \\
Data-based & 0.00 & 0.426 \\
\bottomrule
\end{tabular}
\end{table}
The trajectories of the resulting simulation runs are very similar. The system is driven to the same steady state very consistently. 

Counterintuitively, after approximating both systems with neural networks, the previously easier system (two-stage MSMPR) now requires more computation time for MPC than the previously more difficult system (tubular crystallizer). This fact can be attributed to two main factors. First, the difficulty of the original tubular model stemmed primarily from discretization challenges rather than the inherent complexity of the data. The neural network approximation eliminates this discretization bottleneck. In addition, the two-stage system involves greater data complexity as it requires predicting measurements from both stages, whereas the tubular system only predicts measurements at the outlet. Second, the case studies employ different constraint types: the two-stage system constrains supersaturation in both stages, which is a nonlinear function of the states, while the tubular system constrains temperature, a state variable. Consequently, the MPC problem becomes computationally more demanding for the previously easier two-stage MSMPR system.

\awesomebox[noteblue]{2pt}{\faLightbulb}{noteblue}{
\textbf{\blue{Takeaways}}
\begin{itemize}[leftmargin=*]
    \item \blue{Direct use of first-principle models in model predictive control (MPC) only feasible for MSMPR-type systems with moment-based population balance equation solutions}
    \item \blue{Surrogate-based MPC can match first-principle MPC performance}
    \item \blue{Penalize input changes to prevent surrogate model exploitation by optimizer} 
\end{itemize}
}

\section{Discussion and future perspectives}\label{section:FuturePerspectives}

\begin{figure*}
    \centering
    \includegraphics{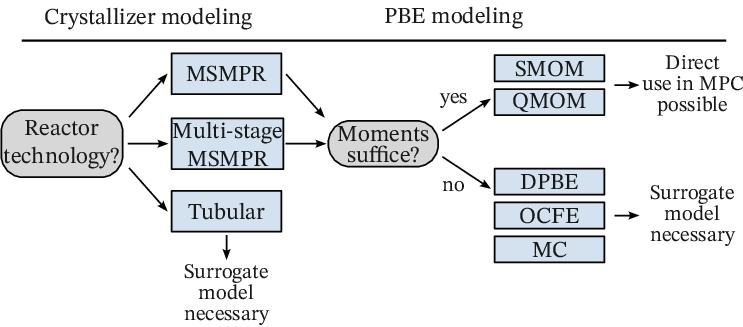}
    \caption{Decision tree showing the necessity of training a surrogate model depending on the modeling choices. A general in-depth discussion for surrogate models can be found in \cite{heReadytouseDeeplearningSurrogate2023}.}
    \label{figure:DecisionTree}
\end{figure*}

Figure~\ref{figure:DecisionTree} presents a decision tree that illustrates when a data-based model may be necessary. In general, it is preferable to avoid the additional step of model approximation because of the limited validity of surrogate models outside of the training data. However, the model used within the MPC controller must be as accurate as possible to ensure effective process control. Therefore, modeling should aim for the highest possible fidelity. In this context, using a surrogate of a detailed model can yield more accurate predictions than relying on an overly simplified model that does not require approximation. For continuous crystallization, the direct use of a first-principle model is only feasible for MSMPR or multi-stage MSMPR systems and only when the PBE is solved via a moment-based method. In all other cases, a surrogate model must be trained, an approach that, as demonstrated in the previous section, can also yield highly accurate results.

For the practitioner, there are two main options for accelerating the use of neural network-based surrogate models in MPC. Either the complexity of the neural network can be reduced or the numerical algorithm to solve the MPC problem can be improved. The latter approach is difficult for the user, as in most cases readily available software tools and solvers are used to solve the MPC problem. In most cases, the only option (without changing the optimization problem) is to select a potentially faster numerical solver. However, the complexity of the neural network can be greatly influenced by the user. Again, in principle, there are two main approaches to reducing the complexity of a standard feedforward neural network. Alternatively, fewer neurons and hidden layers are chosen, or the number of inputs and outputs can be reduced. In general, the number of neurons and hidden layers should always be kept as small as possible but as large as necessary to achieve good approximation accuracy. The number of inputs can be reduced by dimensionality reduction, such as PCA. Reducing the number of outputs has already been discussed in Section~\ref{subsection:SurrogateModelDevelopment} and is achieved by predicting measurements only. Experience shows that the number of outputs and the number of hidden layers have the greatest influence on the computational time of the overall problem (MPC with a neural network as internal model).

\blue{The following subsections discuss directions for extending the methods presented in this tutorial: modeling of novel crystallizers, uncertainty quantification for surrogate models, and control strategies that leverage uncertainty information.}

\subsection{Novel continuous crystallizers}

\blue{The MSMPR and tubular crystallizer models used throughout this tutorial represent idealized systems. In practice, novel continuous crystallizer designs, such as tubular crystallizers with static mixers \cite{alvarezContinuousPlugFlow2010, mathewthomasDesignCharacterizationKenics2022}, continuous oscillatory baffled crystallizers \cite{briggsSeededCrystallizationGlutamic2015,lawtonContinuousCrystallizationPharmaceuticals2009,mcgloneOscillatoryFlowReactors2015,penaProcessIntensificationContinuous2017}, and slug flow crystallizers \cite{termuhlenCharacterizationSlugFormation2019,jiangContinuousFlowTubularCrystallization2014, mozdzierzTunableProteinCrystal2021, rascheMathematicalModelingOptimal2016, terhorstFundamentalsIndustrialCrystallization2015}, have been developed to address limitations in residence time distribution, particle suspension, and fouling. These crystallizers offer operational advantages, but require new modeling efforts to enable model-based control. The core modeling challenge is that novel crystallizers exhibit phenomena that idealized systems cannot capture. Slug flow crystallizers, for example, require tracking individual slugs as they traverse the crystallizer with an accelerating velocity profile \cite{johnsonMultistageModelPredictive2026}. Such models are substantially more complex than the tubular example in this tutorial, making surrogate modeling essential for real-time MPC.}

\blue{Beyond the complexity of individual crystallizer models, process configurations and operating conditions introduce additional challenges. Recycle loops, increasingly used to improve yield, create feedback dynamics that complicate both modeling and control. Fouling affects all crystallizer types but remains difficult to describe from first principles. For fouling, data-based models often become a necessity rather than a convenience. Finally, inline measurement of the particle size distribution remains an open challenge, limiting feedback for closed-loop control in practice \cite{nagyRecentAdvancesMonitoring2013, emmerichOpticalInlineAnalysis2019}. MPC is particularly valuable for these systems because fouling requires periodic shutdowns, making startup optimization a recurring control problem rather than a one-time event.}

\subsection{Parameter estimation and uncertainty quantification}

\blue{Parameter estimation is a major challenge in model development. In particular, for crystallization, the PBE parameters for nucleation, growth, agglomeration, and breakage must be fitted to experimental data \cite{besenhardEvaluationParameterEstimation2015}. In practice, only limited experiments with noisy measurements are available, leading to challenges in parameter identifiability \cite{bosParameterEstimationScientists2007}. The work in \cite{besenhardEvaluationParameterEstimation2015} provides a study on different techniques for identifying parameters within crystallization models using the PBE. The consequence is that model parameters and therefore model predictions remain uncertain.}

\blue{This uncertainty becomes particularly important when using surrogate models for MPC. As discussed previously, surrogate models are only valid within the domain of their training data. Extrapolation produces unreliable predictions, but the controller is unaware of this if the model does not quantify the uncertainty of its predictions. Using uncertainty-aware surrogate models in the controller allows the controller to use the predicted uncertainty and to act on it. The controller can make different decisions depending on whether predictions are trustworthy or not.}

\blue{Several methods exist for training surrogate models with uncertainty quantification. Gaussian processes \cite{rasmussenGaussianProcessesMachine2006} provide well-calibrated uncertainty estimates but scale poorly with large data sets. Bayesian neural networks \cite{jospinHandsonBayesianNeural2020} scale more favorably with large data sets but require slow, approximate inference through sampling or variational methods. Bayesian last layer neural networks \cite{lazaro-gredillaMarginalizedNeuralNetwork2010, watsonLatentDerivativeBayesian2021} offer a practical compromise, enabling a fast analytical evaluation while approximating the full Bayesian inference. Conformalized quantile regression \cite{romanoConformalizedQuantileRegression2019} takes a different approach. By predicting also the quantiles, distribution-free coverage guarantees are achieved.}

\blue{When selecting a method, the distinction between epistemic and aleatoric uncertainty is important \cite{jospinHandsonBayesianNeural2020}. Epistemic uncertainty arises from limited data, for example, uncertain growth rate parameters due to few experiments. Aleatoric uncertainty reflects inherent variability to the system, such as measurement noise or stochastic nucleation events. The practical implication is that epistemic uncertainty can be reduced by collecting more data, whereas aleatoric uncertainty cannot be reduced. Understanding this distinction helps practitioners decide whether additional experiments would improve model reliability or whether the irreducible variance must instead be handled through robust control design.}

\subsection{Robust and stochastic model predictive control}

After developing a model that produces uncertain predictions, whether through uncertain parameters or directly via a surrogate model, it becomes essential to effectively incorporate this uncertainty into the control algorithm. The two most popular directions are robust MPC and stochastic MPC. In robust MPC, a disturbance or an uncertain parameter is assumed to belong to a bounded set. Subsequently, constraint satisfaction should be guaranteed for all possible values of the disturbance \cite{mayneModelPredictiveControl2014}. Stochastic MPC assumes that the disturbance or uncertainty is stochastic. Constraints are satisfied with a certain probability. An overview of stochastic MPC can be found in \cite{mesbahStochasticModelPredictive2016}.
 
The trade-off between control objectives and constraint satisfaction is inherent in these methods. The development of new methods is mainly concerned with improving computational efficiency and deriving theoretical guarantees, especially for nonlinear models. \blue{The theoretical foundations for these methods are well established and progress in uncertainty quantification, as discussed above, directly enables their practical implementation.} However, more advanced methods have not been experimentally demonstrated in continuous crystallization applications. \blue{The main barrier is not the lack of methodology but the lack of conventional MPC implementations to continuous crystallization. Conventional MPC itself is not yet widely applied in crystallization. As machine learning-based surrogate models and faster computing make standard MPC more accessible, robust and stochastic extensions become the natural next step.} We believe that the tools presented in this tutorial can help in making advanced model-based robust control of continuous crystallization a reality. 

\blue{Despite these developments, we acknowledge that experimental implementation remains limited. Most applications of MPC with machine learning models have only been demonstrated in simulation \cite{limaApplicationsMachineLearning2025}, and PID control still dominates industrial crystallization \cite{orehekContinuousCrystallizationProcesses2021}. The transition from batch to continuous processing, particularly in the pharmaceutical industry, is further slowed down by strict regulatory requirements \cite{ardenIndustry40Pharmaceutical2021}. We expect that as the awareness of the potential of advance control increases,  surrogate modeling tools mature and standard MPC becomes more accessible, these barriers will diminish.}




\section{Conclusion}\label{section:Conclusion}

\blue{The application of MPC to crystallization has traditionally been limited to well-mixed systems with simplified representations of the particle size distribution. Recent advances in surrogate modeling are now making it feasible to extend MPC to spatially distributed crystallizers while retaining full particle size information. This tutorial guides practitioners through the complete workflow enabling MPC for spatially distributed crystallizers with full particle size information. We first present how to obtain accurate models through PBE solution methods and crystallizer modeling. We then show how to transform these models into computationally efficient surrogates suitable for optimization-based control. Finally, we demonstrate how to deploy these surrogates within MPC frameworks. The two case studies presented in this tutorial illustrate how these components integrate in practice. The well-mixed MSMPR system demonstrates that surrogate-based MPC achieves performance comparable to optimization with the original first-principle model. The spatially distributed slug-flow crystallizer confirms that the methodology extends to more complex configurations, where direct optimization would be intractable. Several challenges remain. Surrogate models are only valid for interpolation within their training domain, which limits applicability when operating conditions shift. Most applications have only been demonstrated in simulation, and experimental validation on real continuous crystallizers remains scarce. Promising directions include uncertainty quantification methods, such as Bayesian approaches and conformalized quantile regression, that can make surrogate predictions more trustworthy and enable robust or stochastic MPC formulations. As surrogate modeling makes standard MPC more accessible for continuous crystallization, these extensions become a natural next step. Novel crystallizer designs will require further modeling advances, while broader experimental validation on industrial systems remains essential to translate these methods into practice.}

\section*{Declaration of Generative AI and AI-assisted technologies in the writing process}

During the preparation of this work the authors used Claude Sonnet 4 for grammar checks and slight reformulations. After using this tool, the authors reviewed and edited the content as needed and take full responsibility for the content of the published article.

\printcredits

\bibliographystyle{model1-num-names}

\bibliography{CrystTutorial}



\end{document}